\documentclass[aps,prd,twocolumn,
superscriptaddress,nofootinbib,amsfonts,amssymb,amsmath,longbibliography]{revtex4-1}
\usepackage{graphicx,bm,color,wrapfig,cancel} 
\usepackage[colorlinks,linkcolor=red,anchorcolor=blue,citecolor=green]{hyperref}
\usepackage[normalem]{ulem}
\usepackage[utf8]{inputenc}

\begin{document}

\title{Hunting dark energy with pressure-dependent photon-photon scattering \\
--- {\it Dedicated to the late professor Yasunori Fujii} ---}

\author{Taishi Katsuragawa}
\email{taishi@mail.ccnu.edu.cn}
\affiliation{Institute of Astrophysics, Central China Normal University, Wuhan 430079, China}
\author{Shinya Matsuzaki}
\email{synya@jlu.edu.cn}
\affiliation{Center for Theoretical Physics and College of Physics, Jilin University, Changchun, 130012, China}
\author{Kensuke Homma}
\email{khomma@hiroshima-u.ac.jp}
\affiliation{Graduate School of Advanced Science and Engineering, Hiroshima University, Kagamiyama, Higashi-Hiroshima, Hiroshima 739-8526, Japan}

\begin{abstract}
Toward understanding of dark energy, we propose a novel method to directly produce a chameleon particle and force its decay under controlled gas pressure in a laboratory-based experiment. 
{\it Chameleon gravity}, characterized by its varying mass depending on its environment, could be a source of dark energy, which is predicted in modified gravity. 
A remarkable finding is a correspondence between the varying mass and a characteristic pressure dependence of a stimulated photon-photon scattering rate in a dilute gas surrounding a focused photon-beam spot. 
By observing a steep pressure dependence in the scattering rate, we can directly extract the characteristic feature of the chameleon mechanism.
As a benchmark model of modified gravity consistent with the present cosmological observations, a reduced $F(R)$ gravity is introduced in the laboratory scale.
We then demonstrate that the proposed method indeed enables a wide-ranging parameter scan of such a chameleon model with the varying mass around $(0.1-1)~[\mu \mathrm{eV}]$ by controlling pressure values. 
\end{abstract}

\maketitle

\section{Introduction}
A variety of independent observations have confirmed the accelerated expansion of the Universe,
which indicates an unknown energy called dark energy (DE).
Modified gravity theory has been considered one of the solutions to the DE problem
as an alternative to the {\it ad hoc} introduction of the cosmological constant to general relativity. 
Among the modified gravity theories, scalar-tensor theory introduces a new scalar field~\cite{Caldwell:1997ii,Peebles:2002gy,Copeland:2006wr}. 
This scalar field plays a role of dynamical DE,
and its coupling to ordinary matter induces the so-called fifth force.
The phenomenology of the new scalar field and the induced fifth force have been developed to constrain modified gravity:
possible deviations from the gravitational inverse-square law~\cite{Will:2014kxa,Adelberger:2003zx,Spero:1980zz,Kapner:2006si}
and time-varying parameters of the standard model of particle physics~\cite{Marion:2002iw,Peik:2004qn,Damour:1996zw,Srianand:2004mq}.

It is known that a dilatonic scalar field shows up in modified gravity for DE 
through the Weyl transformation of the metric~\cite{DeFelice:2010aj,Nojiri:2010wj,Clifton:2011jh,Capozziello:2011et,Nojiri:2017ncd}.
A class of such a dilaton is called chameleon field, 
which is named after a remarkable feature of the environment dependence, the chameleon mechanism~\cite{Khoury:2003aq}.
The coupling to matter influences the effective mass of the chameleon field,
which is analogous to an in-medium mass (an effective mass) in heavy fermion materials.
Consequently, the chameleon mechanism screens the fifth force in a high-density environment, 
which allows the modified gravity to be consistent with short-scale local experiments and observations.

\begin{figure}[htbp]
\centering
\includegraphics[width=0.4\textwidth]{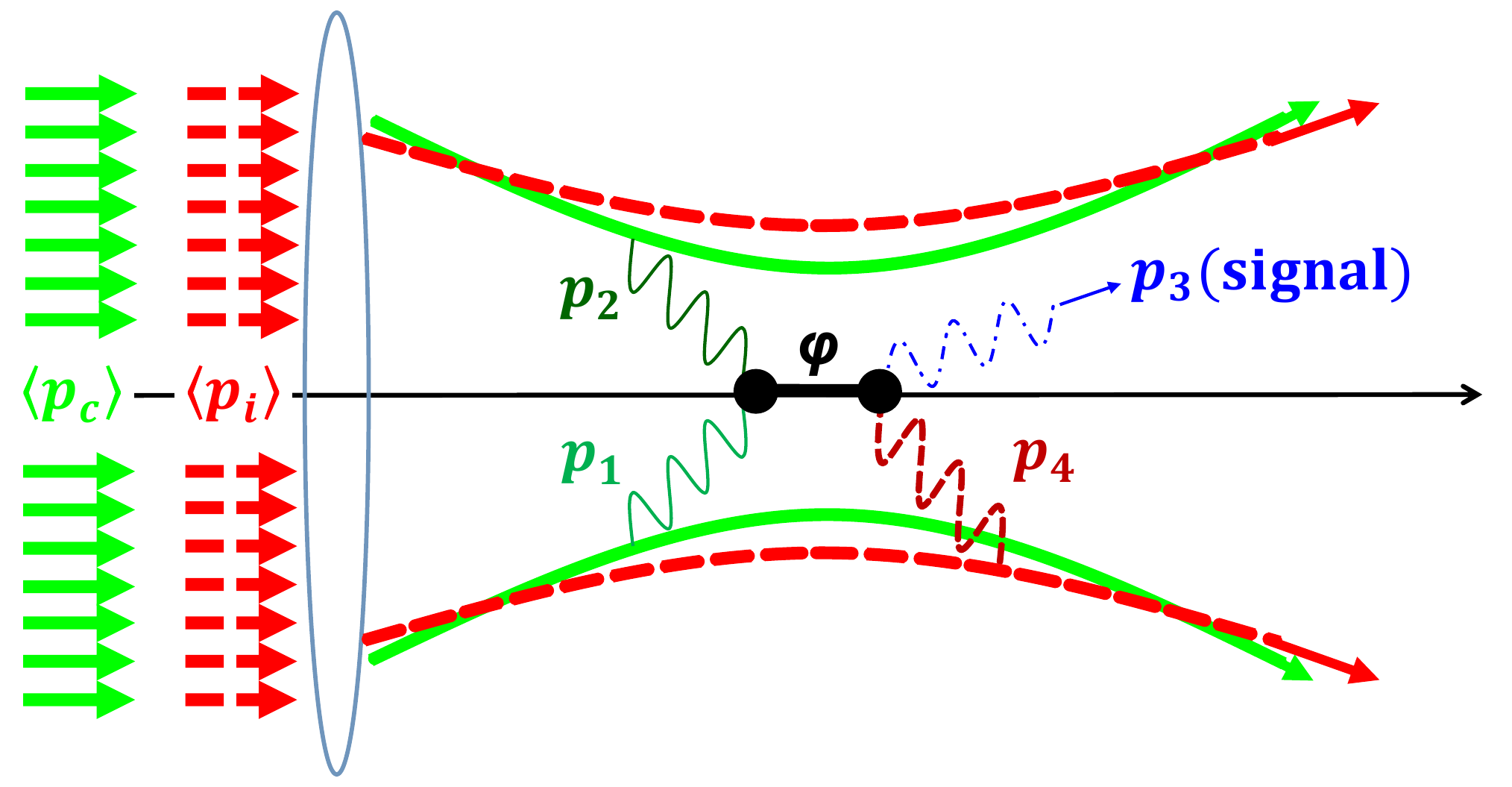}
\caption{
Concept of stimulated resonant photon-photon scattering in quasiparallel collision geometry.
}
\label{Fig: concept}
\end{figure}

There have been attempts to search for the chameleon field using nongravitational experiments~\cite{Burrage:2017qrf,Chou:2008gr,Steffen:2010ze,Rybka:2010ah,Brax:2009aw,CAST:2018bce,Levshakov:2009ft,Vagnozzi:2021quy}.
Recently, it was suggested that a stimulated radar-collider concept may be applicable to probe gravitationally coupled scalar fields~\cite{Homma:2019rqb}, as illustrated in Fig.~\ref{Fig: concept}.
The method resonantly produces a low-mass scalar field $\varphi$ in quasiparallel photon-photon scattering and simultaneously stimulates its decay by combining two different frequency photon beams along the same optical axis and focusing them into a vacuum. As a consequence of the stimulated resonant scattering, a frequency-shifted photon can be generated as a clear signal of scattering.
This provides the possibility of directly exploring the microscopic nature relevant to DE through the photon-photon scattering experiment,
despite the fact that the coupling strength considered to be of the order of (or even less than) the gravitational constant.

We find a new and remarkable potential to probe the chameleon particle and the chameleon mechanism 
in the stimulated photon-photon scattering experiment in~\cite{Homma:2019rqb}.
It is possible to directly extract the chameleon mechanism via a 
correspondence between the varying mass and a characteristic pressure dependence of the 
stimulated photon-photon scattering rate in a dilute gas surrounding a focused photon-beam spot. 
The methodology is applied to test  
a class of chameleon models predicted in the viable DE models of $F(R)$ gravity~\cite{Nunes:2016drj}. 
We discuss the pressure dependence of the signal in viable parameter spaces of 
the Starobinsky model~\cite{Starobinsky:2007hu} and the Hu-Sawicki model~\cite{Hu:2007nk}, which are widely used in late-time cosmology. 
It will be shown that 
the pressure dependence can be observed as a signal photon number with a tenth power steepness,  
which is clearly separable from the background atomic process.

\section{Extendable chameleon mass: coupling domain}
We begin with the theoretical basis on the chameleon field and show the expected sensitivity to probe the chameleon field using the experimental idea given in~\cite{Homma:2019rqb} 
projected onto the mass-coupling  space. 
A chameleon field $\phi$ coupling to matter fields is described in scalar-tensor theory as
\begin{align}
S =& \int d^{4}x\sqrt{-g_{*}}
\left[ \frac{M^{2}_{\mathrm{pl}}}{2} R_{*}
-\frac{1}{2} g_{*}^{\mu\nu}(\partial_{\mu}\varphi)(\partial_{\nu}\varphi)-V(\varphi)\right]
\nonumber \\
& \qquad 
+ \int d^{4}x\sqrt{-g_{*}} \mathcal{L}_{\mathrm{M}}\left[ g^{\mu\nu}_{*}, \varphi, \Phi_{i} \right]
\label{Eq: Action in Einstein frame}
\, , 
\end{align}
where $\mathcal{L}_{\mathrm{M}}$ is the Lagrangian density of arbitrary matter fields $\Phi_{i}$ 
and $M_{\mathrm{pl}}$ is the reduced Planck mass, $M_{\mathrm{pl}} \simeq 2.44 \times10^{27}~[\mathrm{eV}]$.
The Einstein frame metric $g_{*\mu\nu}$ and the Jordan frame metric $g_{\mu\nu}$ are related to each other via  
$g_{* \mu\nu}=\mathrm{e}^{2\beta_{i}\varphi/M_{\mathrm{pl}}}g_{\mu\nu}$.
In general, scalar-tensor theory allows us to parametrize the coupling constants $\beta_{i}$
between the chameleon $\varphi$ and other fields $\Phi_{i}$.
We conventionally define the effective potential of the chameleon field as follows:
\begin{align}
V_\mathrm{eff}(\varphi)\equiv V(\varphi)  - \frac{1}{4} \sum_{i} e^{-\frac{4\beta_{i}\varphi}{M_{\mathrm{pl}}}} T^{\mu}_{\ \mu (i)}  
\, ,
\label{Eq: Effective potential}
\end{align} 
where $T_{\mu \nu (i)}$ is the energy-momentum tensor of $\Phi_{i}$ in the Jordan frame.~\footnote{
Here the $\varphi$ dependence in $T^\mu_{\ \mu}$ has been disregarded, which suffices for this study of a static environmental effect 
(a residual gas pressure in an experimental chamber) constructed only from matter, external to the chameleon dynamics. 
A similar prescription has been applied in a different context in the literature~\cite{Katsuragawa:2018wbe}.  
}
Because of the dilatonic coupling between $\varphi$ and $T^{\mu}_{\ \mu}$, 
the effective potential and the chameleon mass acquire the environment dependence which drives the chameleon mechanism.

The chameleon can interact with the electromagnetic field through the quantum trace anomaly, which generates the coupling to two photons. 
This coupling generation can be viewed through the anomalous Weyl transformation~\cite{Fujikawa:1980vr, Fujii:2015iuz, Ferreira:2016kxi, Katsuragawa:2016yir}, 
and the derived interaction terms can be cast in the following form: 
\begin{align}
\begin{split}
\mathcal{L}_\mathrm{M}
&= \frac{1}{4} e^{- \frac{4\beta_{\gamma}\varphi}{M_{\mathrm{pl}}}} T^{\mu}_{\ \mu (\mathrm{anomaly})} 
\\ 
& = 
- \frac{1}{4} e^{- \frac{4\beta_{\gamma}\varphi}{M_{\mathrm{pl}}}}
 \left( 
 \frac{b_{\mathrm{em}} \cdot \alpha_{\rm em}}{8 \pi} F_{\mu \nu} F^{\mu \nu} 
\right) 
\, .
\end{split}
\label{Eq: chameleon photon interaction}
\end{align}
Here $\alpha_{\rm em}$ 
is the fine-structure constant of the electromagnetic coupling and $b_{\rm em}$ denotes a beta function coefficient
which would include all the charged particle contributions at the loop level of the standard model of particle physics: 
$b_{\rm em} = 11/3$ at the one-loop level. 
The relevant interaction between the chameleon and the photons is thus written as~\footnote{
Here we ignore the frame difference between the Jordan and Einstein frames, 
which arises as $e^{\kappa \varphi_{\min}}$, 
with the potential minimum $\kappa \varphi_{\min}$ assumed to be small~\cite{Katsuragawa:2016yir}.
}
\begin{align}
\mathcal{L}_\mathrm{\varphi \gamma \gamma} 
&= \frac{1}{4} \frac{\tilde{\beta}_{\gamma}\varphi}{M_{\mathrm{pl}}}  F_{\mu \nu} F^{\mu \nu} 
\, , \ 
\tilde{\beta}_{\gamma} \equiv \frac{b_{\rm em} \cdot \alpha_{\mathrm{em}}}{2\pi}\beta_{\gamma}
\label{Eq: photon coupling}
\, .
\end{align}
Furthermore, various matter consisting of an experimental environment may also contribute to the chameleon mechanism.
The matter contribution is cast into the matter energy density $T^{\mu}_{\mu} \approx - \rho_{\mathrm{m}}$,
which is evaluated in each chameleon-search experiment.

\begin{figure}[htbp]
\centering
\includegraphics[width=0.45\textwidth]{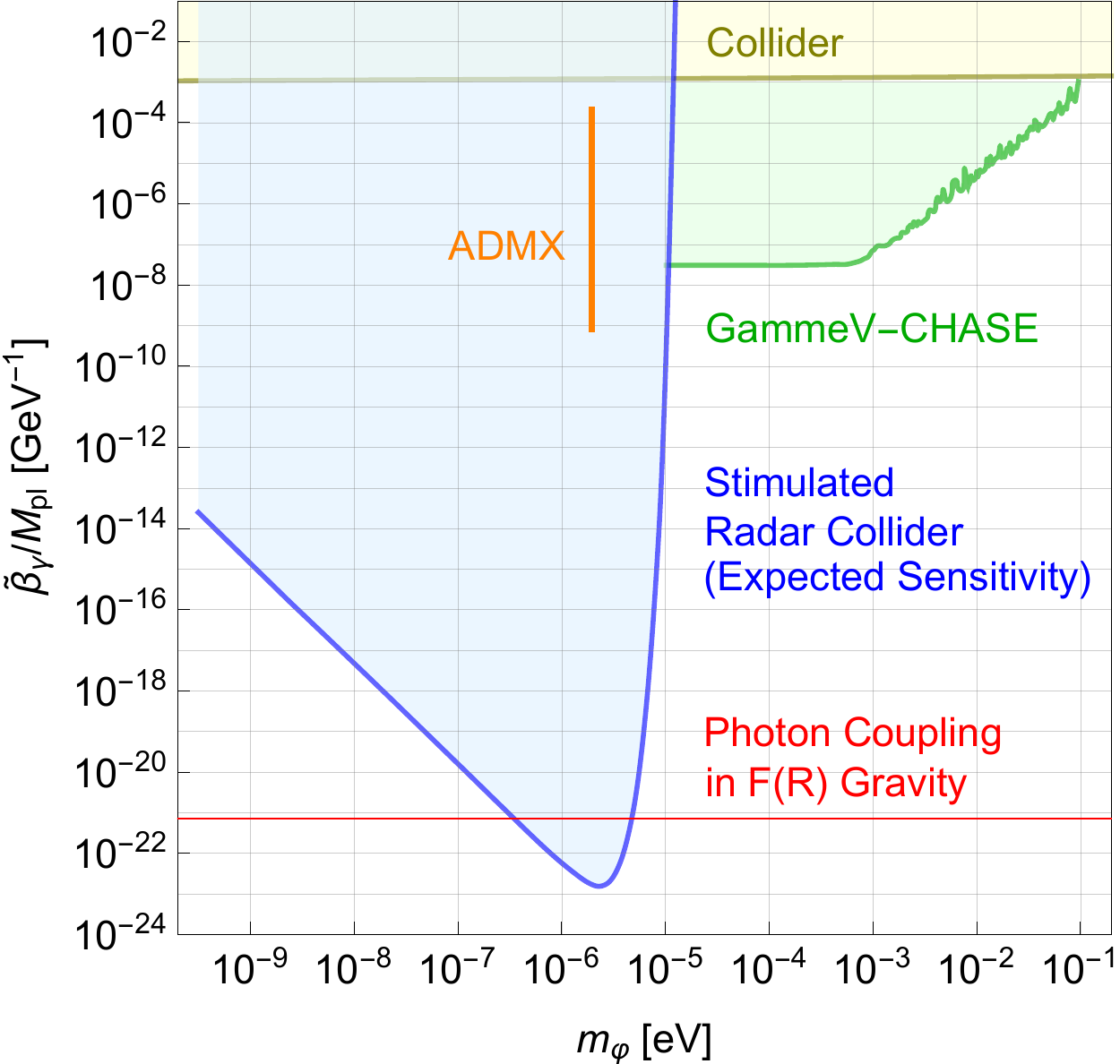}
\caption{
Expected detection sensitivity of the chameleon field in the presently proposed photon-photon scattering experiment (blue region) 
by focusing two pulses with $100~[\mathrm{J/ns}]$ consisting of gigahertz-band photons in the quasiparallel geometry evaluated in Ref.~\cite{Homma:2019rqb}. 
The red horizontal line corresponds to the chameleon-photon coupling $\tilde{\beta}_{\gamma} = \frac{b_{\rm em} \alpha_{\rm em}}{2 \sqrt{6} \pi}$ when we take $\beta_{\gamma}=\frac{1}{\sqrt{6}}$ in Eq.~\eqref{Eq: photon coupling}.
The chameleon in $F(R)$ gravity is hunted along the red line that overlaps with the detection sensitivity domain. 
Constraints from GammeV-CHASE (green region), ADMX (orange line), and collider experiments (yellow region) are shown for comparison. 
}
\label{Fig: sensitivity}
\end{figure}

Based on the following parametrization,~\footnote{
The sign in the exponent follows from the definition of the Weyl transformation and the chameleon field.
}
\begin{align}
V_\mathrm{eff}(\varphi) =  
V(\varphi) + \frac{1}{4} F_{\mu \nu}F^{\mu \nu}  e^{\frac{\tilde{\beta}_{\gamma}\varphi}{M_{\mathrm{pl}}}}
+ \rho_{\mathrm{m}} e^{\frac{\beta_{\mathrm{m}}\varphi}{M_{\mathrm{pl}}}}
\label{Eq: parametrization}
\, ,
\end{align}
the nongravitational experiments placed the limit on $\tilde{\beta}_{\gamma}$~\cite{Burrage:2017qrf}.
In Fig.~\ref{Fig: sensitivity} we summarize current constraints 
and the expected detection sensitivity in the stimulated radar collider~\cite{Homma:2019rqb}
in the mass-coupling parameter space.~\footnote{
To evaluate the coupling-mass relation, Eq.~(3.1) in Ref.~\cite{Homma:2019rqb} was used 
by substituting an assumed number of frequency-shifted photons, $N_{obs} = 100$, as summarized in Table 1 of Ref.~\cite{Homma:2019rqb}, 
and we numerically solve the equation to get the coupling strength $g/M$ as a function of mass. 
The core formula for the stimulated signal yield, ${\cal }Y_{c+i}$ in Eq.~(3.1), is derived as Eq.~(A.54) in the Appendix. 
Since the generic effective interaction Lagrangian in Eq.~(1.1) of Ref.~\cite{Homma:2019rqb} is exactly the same as Eq.~\eqref{Eq: photon coupling}
when the coupling is redefined as $g/M = -\tilde{\beta}_{\gamma}/M_{pl}$, 
once the coupling strength is determined with the same set of the experimental parameters as a function of mass, 
we have only to relabel the coupling strength in Fig.~\ref{Fig: sensitivity}.
}
GammeV CHameleon Afterglow SEarch (GammeV-CHASE) constrains $\tilde{\beta}_{\gamma} \lesssim 7.1 \times 10^{10}$ 
when $m_{\varphi}\lesssim\mathcal{O}(10^{-3})~ [\mathrm{eV}]$~\cite{Steffen:2010ze}.
In terms of the beyond-the-standard-model (BSM) particle-search experiments, 
the Axion Dark Matter Experiment (ADMX) excluded $2\times10^{9} \lesssim \tilde{\beta}_{\gamma} \lesssim 5\times10^{14}$ 
for $m_{\varphi} \sim 1.95~[\mathrm{\mu eV}]$~\cite{Rybka:2010ah},
and the collider experiments placed the limit 
$\tilde{\beta}_{\gamma} \lesssim 2\times10^{15}$ for $10^{-11}~[\mathrm{eV}] \lesssim m_{\varphi} \lesssim 10^{4}~[\mathrm{eV}]$~\cite{Brax:2009aw}. 
We note that  
the CERN Axion Solar Telescope has placed the limit on the coupling space ($\beta_{m}$, $\tilde{\beta}_{\gamma}$) as  
$\tilde{\beta}_{\gamma} \lesssim 5.7\times10^{10}$ for $1<\beta_{m}<10^{6}$~\cite{CAST:2018bce}.
As evident in Fig.~\ref{Fig: sensitivity}, 
the simulated radar-collider experiment in~\cite{Homma:2019rqb} can improve the current constraints on $\tilde{\beta}_{\gamma}$ 
in terms of scalar-tensor theory by more than the tenth order of magnitude.~\footnote{ 
The method depicted in Fig.~\ref{Fig: concept} is based on the following stimulated two-body photon-photon scattering process:
$ \langle p_c (p_1) \rangle + \langle p_c (p_2) \rangle \rightarrow  p_3 + \langle p_i (p_4) \rangle $, 
where $\langle (\cdot) \rangle$ indicates that two incident photons with four-momenta $p_1$ and $p_2$ stochastically annihilate from a broadband coherent creation beam with its central four-momentum $p_c$,
while $p_4$ is created as a broadband, copropagating inducing coherent beam with the central four-momentum $p_i$. 
This process yields the stimulated emission of a $p_3$ signal photon via induced decay of a produced resonance state through 
energy-momentum conservation with $p_4$.
To introduce the comoving inducing beam technically, 
$\langle p_c \rangle$ and $\langle p_i \rangle$ beams are combined along a common optical axis and simultaneously focused into vacuum~\cite{Homma:2014rja}. 
Thus, this process looks as if a frequency-shifted photon is generated from vacuum by mixing two-color beams.
This will be the first experiment to explore DE gravitationally coupled to photon with $\tilde{\beta}_\gamma \lesssim \mathcal{O}(1)$,
which the existing chameleon search experiments cannot reach.
}

\section{Chameleon in $F(R)$ gravity}
When one takes full advantage of the accessibility to $\tilde{\beta}_\gamma \lesssim \mathcal{O}(1)$,
$F(R)$ gravity can provide a suitable benchmark model of the chameleon.
$F(R)$ stands for a function of the Ricci scalar $R$,
and the Einstein-Hilbert action of the general relativity is replaced by $F(R)$.
By the Weyl transformation defined as 
$\mathrm{e}^{2\sqrt{1/6}\varphi/M_{\mathrm{pl}}}\equiv F_{R}(R)$,
$F(R)$ gravity turns into a specific case of scalar-tensor theory [Eq.~\eqref{Eq: Action in Einstein frame}].
Here the subscript $R$ for $F(R)$ denotes the derivative with respect to $R$.
The potential $V(\varphi)$ is written as a function of the Ricci scalar,
\begin{align}
V(\varphi) \equiv \frac{M^{2}_{\mathrm{pl}}}{2}\frac{R(\varphi)F_{R}(R(\varphi))-F(R(\varphi))}{F_{R}^{2}(R(\varphi))}
\label{Eq: original potential}
\, .
\end{align}

Unlike in generic scalar-tensor theory,
the coupling constant $\beta_{i}$ in Eq.~\eqref{Eq: Effective potential} is not parametric, but rather fixed as $\beta_{i}=1/\sqrt{6}$ for all matter fields in $F(R)$ gravity~\cite{Maeda:1988ab}.
In terms of Eq.~\eqref{Eq: parametrization},
$F(R)$ gravity gives $\tilde{\beta}_{\gamma} =\mathcal{O}(10^{-3})$ and $\beta_{\mathrm{m}} = \mathcal{O}(1)$,
which eludes those in the existing nongravitational tests.
Because the coupling constants are independent of $F(R)$ function,
$F(R)$ gravity models are testable if they predict chameleon fields with masses around $m_{\varphi}=(0.1-1)~[\mathrm{\mu eV}]$ as in Fig~\ref{Fig: sensitivity}.

\section{Matter sources for chameleon effects}
In Eq.~\eqref{Eq: Effective potential}, 
we have mainly two ambient sources, photon and gas densities, 
in scanning the effective mass of the chameleon,
\begin{align}
T^{\mu}_{\ \mu} = T^{\mu}_{\ \mu (\mathrm{photon})}  + T^{\mu}_{\ \mu (\mathrm{gas})} 
\, .
\end{align}
We evaluate those two ingredients in $T^{\mu}_{\ \mu}$ 
in light of the proposed experimental setup. 
The photon contribution in Eq.~\eqref{Eq: photon coupling} goes like 
$T^{\mu}_{\ \mu (\mathrm{photon})} \propto F_{\mu \nu} F^{\mu \nu} 
= 2(\mathbf{B}^{2} - \mathbf{E}^{2})$, 
with the electric field $\mathbf{E}$ and magnetic field $\mathbf{B}$. 
This contribution vanishes 
in the case with the ordinary plane electromagnetic wave where 
$\mathbf{E}$ and $\mathbf{B}$ have the same amplitude,
while in the case of the focused geometry in Fig.\ref{Fig: concept} amplitudes of the electric and magnetic fields are given by nontrivial spatial distribution functions (as is explained in the Appendix A~\footnote{
The contents for the corresponding part in the Appendix A are based entirely on an unpublished doctoral thesis by Yuichiro Monden, 
which was also the grounds of Refs.~\cite{PhysRevLett.107.073602,PhysRevA.86.033810}. 
}). 
It turns out, however, that the square of the field strength vanishes at each point on the focal plane in the case of circularly polarized beams~\cite{wolf1959electromagnetic, richards1959electromagnetic, stamnes2017waves}.
Therefore, with the circularly polarized beams proposed in~\cite{Homma:2019rqb}, we can ignore the contribution from the background photon density.~\footnote{ 
In Ref.~\cite{Chou:2008gr} 
the photon contribution was simply assumed to be of the same order as the gas contribution. 
}

Regarding the gas contribution, 
we assume the conventional perfect-fluid description because of the difficulties in treating the atoms or molecules using the fundamental field prescription.
Then the trace of the energy-momentum tensor is evaluated as 
$T^{\mu}_{\ \mu (\mathrm{gas})}  = - (\rho - 3P)$,
where $\rho$ and $P$ represent the density and pressure, respectively.
Gas in an experimental chamber exhibits a low pressure ($P \ll \rho$) and a high temperature; 
hence, we assume  the equation of state for the ideal gas. 
In that case the trace of the energy-momentum tensor can be approximated 
as follows: 
\begin{align}
- T^{\mu}_{\ \mu (\mathrm{gas})} \approx \rho~[\mathrm{kg/m^{3}}] = \frac{P~[\mathrm{Pa}]}{\mathcal{R}T} \,, 
\label{Eq: energy density}
\end{align}
where $T~[\mathrm{K}]$ is the temperature and $\mathcal{R}$ is the specific gas constant. 
$\mathcal{R}$ is defined as the molar gas constant $\mathcal{R}_{u}$ divided by the molar mass of the gas $M~[\mathrm{kg/mol}]$,
$\mathcal{R}=\mathcal{R}_{u}/M$, 
where $\mathcal{R}_{u}=8.31446~[\mathrm{m^{3}\cdot Pa \cdot K^{-1} \cdot mol^{-1}}]$.

In the proposed setup we assume an ultrahigh vacuum chamber to reduce background frequency-shifted photons caused by the residual atoms, and such a low-density environment also weakens the chameleon mechanism.
However, because the gas density is controllable by the gas pressure, as in Eq.~\eqref{Eq: energy density}, over several orders of magnitude, 
in principle, we can measure how the signal is weakened as a function of the gas pressure, which is the characteristic feature of the chameleon mechanism.

\section{Reduced $F(R)$ form for modeling DE}
As one of cosmological models in $F(R)$ gravity, let us consider the $R^{2}$-corrected DE models~\cite{Appleby:2009uf}:
\begin{align}
    F(R) = R + f_{\mathrm{DE}}(R) + \alpha R^{2} 
    \label{Eq: F(R) model0}
\end{align}
where $\alpha$ is a positive parameter.
The second term $f_{\mathrm{DE}}(R)$ represents the modification for DE,
and the third term $\alpha R^{2}$ cures the singularity problem in generic $f_{\mathrm{DE}}(R)$ models~\cite{Frolov:2008uf,Nojiri:2008fk}.
As representatives of $f_{\mathrm{DE}}(R)$, 
we consider the Starobinsky model~\cite{Starobinsky:2007hu} and the Hu-Sawicki model~\cite{Hu:2007nk}, which have been widely used in the study of late-time cosmology.

By considering a fact that the typical energy scale is higher than the DE scale in the laboratory environment,
it is unnecessary to adopt the exact form of $f_{\mathrm{DE}}(R)$.
The large-curvature limit $R/R_{c} \gg 1$, where $R_{c}$ corresponds to the current vacuum curvature, 
is a good enough approximation for local experiments.
In the large-curvature limit, the above two models are reduced to the same form (as explained in Appendix B),
and Eq.~\eqref{Eq: F(R) model0} approximates to the following:
\begin{align}
F(R) = R - \lambda R_{c} \left[ 1 - \left( \frac{R}{R_{c}} \right)^{-2n} \right] + \alpha R^{2} 
\label{Eq: F(R) model}
\, ,
\end{align}
where $\lambda$ and $n$ are positive parameters.
The coefficient $\lambda R_c$ in the second term can be considered the cosmological constant $\lambda R_{c} = 2 \Lambda$,
where $\Lambda\simeq 4.2 \times10^{-66}\ [\mathrm{eV}^{2}]$~\cite{Aghanim:2018eyx}.
In this paper, we adopt Eq.~\eqref{Eq: F(R) model} to investigate the $R^{2}$-corrected Starobinsky and Hu-Sawicki models
in a laboratory-based experiment,
thereby respecting a parameter space $(\lambda, n)$ that is consistent with the cosmological observations.

\section{Chameleon mass dependence on the reduced $F(R)$ parameters}
Next, we study the chameleon mass under the background gas pressure surrounding the scattering point, equivalently, the gas density with a fixed volume in the experimental chamber.
The second derivative of the effective potential at the potential minimum gives the chameleon  mass $m_{\varphi}$.
Using Eq.~\eqref{Eq: original potential}, we can write the chameleon mass formula in terms of the $F(R)$ function and its derivatives:
\begin{align}
\begin{split}
m_{\varphi}^{2} 
&= V(\varphi_{\min})_{\mathrm{eff}, \varphi \varphi}
\\
&= \frac{1}{3F_{RR}(R_{\min})} \left[ 1 - \frac{R_{\min} F_{RR} (R_{\min}) }{F_{R}(R_{\min})} \right]
\end{split}
\label{Eq: mass formula}
\end{align}
where $R_{\min}$ is determined by the stationary condition $V(\varphi_{\min})_{\mathrm{eff}, \varphi}=0$
given by $2F(R_{\min}) - R_{\min} F_{R}(R_{\min}) + \kappa^{2} T^{\mu}_{\mu} = 0$.
In the limit where $R/R_{c} \gg 1$, the stationary condition gives $R_{\min} \approx -\kappa^{2} T^{\mu}_{\mu}$ in the model [Eq.~\eqref{Eq: F(R) model}].
Using the relation $2\Lambda=\lambda R_{c}$, 
we obtain the chameleon mass in the analytic form as~\cite{Katsuragawa:2017wge}
\begin{align}
m_{\varphi}^{2}
&= \frac{\Lambda}{12n(2n+1) \left(\frac{\lambda}{2} \right)^{-2n} \left(-\frac{T^{\mu}_{\mu}}{\rho_{\Lambda}}\right)^{-2(n+1)} + 6\alpha \Lambda}
\label{Eq: chameleon mass formula}
\, ,
\end{align}
where we have defined the DE density as $\rho_{\Lambda} \equiv \Lambda/\kappa^{2}$
and taken the limit $\Lambda \ll R \ll 1/\alpha$.

\begin{figure}[htbp]
\centering
\includegraphics[width=0.45\textwidth]{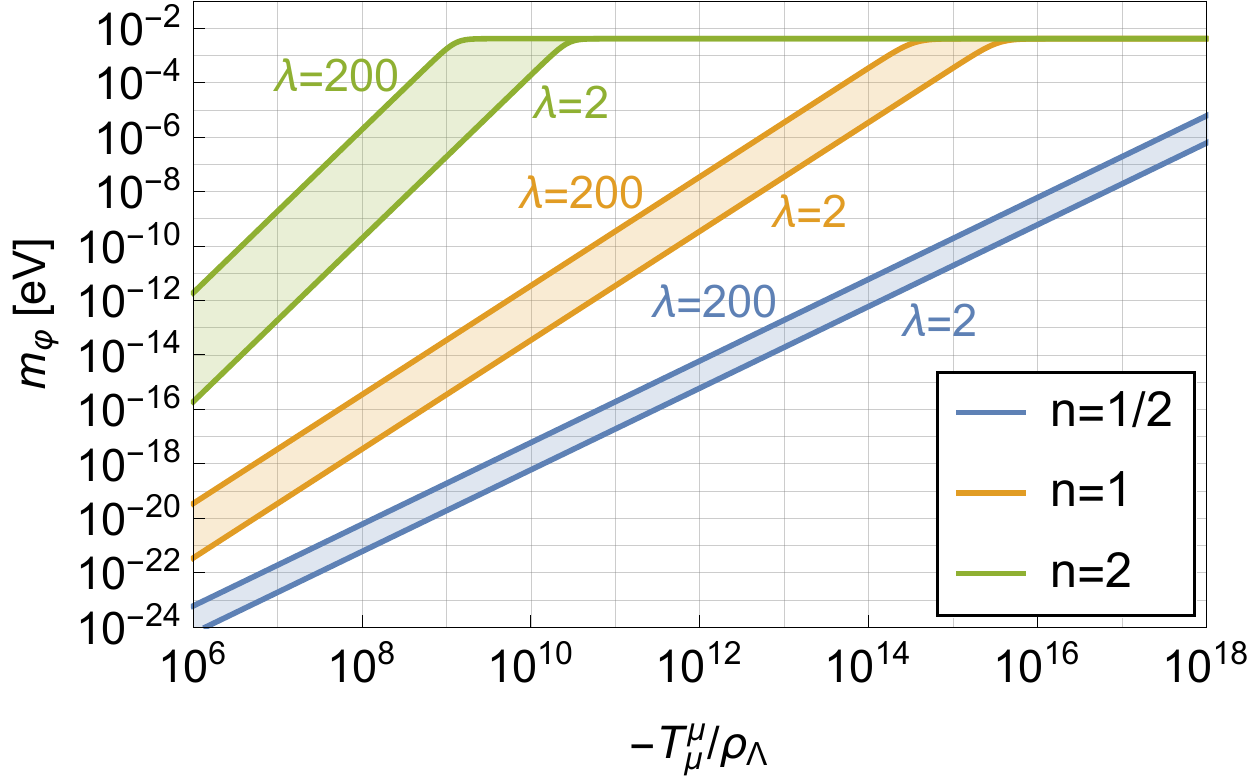}
\caption{
Chameleon mass vs energy-momentum tensor in Eq.~\eqref{Eq: chameleon mass formula}
for $n=1/2, 1, 2$ with $\Lambda =4.2 \times10^{-66}\ [\mathrm{eV}^{2}]$.
For each $n$, $\lambda=2$ and $\lambda=200$ correspond to left and right solid lines, respectively.
As a referenced value, 
we have chosen $\alpha = 10^{4}~[\mathrm{eV^{-2}}]$ to be consistent with the current constraint from the fifth-force experiment,
$\alpha \leq 2.3 \times 10^{4}~[\mathrm{eV^{-2}}]$~\cite{Cembranos:2008gj}. 
}
\label{Fig: chameleon mass plot}
\end{figure}

Figure~\ref{Fig: chameleon mass plot} shows the chameleon mass given in Eq.~\eqref{Eq: chameleon mass formula} as a function of $T^{\mu}_{\ \mu}$
when one varies $\lambda$ from $2$ to $200$.
This range of $\lambda$ overlaps with the cosmological constraint at the $2\sigma$ confidence level~\cite{Nunes:2016drj} 
for the Starobinsky model with $n=1$ and for the Hu-Sawicki model with $n=1/2$.
For a general $n$, a larger $\lambda$ may be allowed 
because $\lambda \rightarrow \infty$ at $\lambda R_{c} = 2\Lambda$ corresponds to the $\Lambda$ cold dark matter model.

We apply the model to a low pressure interval $10^{-8}~[\mathrm{Pa}] <P<10^{-6}~[\mathrm{Pa}]$ within the reach of the current vacuum technology.
In such an ultrahigh vacuum chamber, the residual gas consists mainly of hydrogen molecules,
and Eq.~\eqref{Eq: energy density} for the hydrogen at $T = 300~[\mathrm{K}]$ gives
\begin{align}
1.4\times10^{12} &\lesssim - \frac{T^{\mu}_{\mu}}{\rho_{\Lambda}} \lesssim 1.4 \times 10^{14} 
\, 
\end{align}
with the observed DE density $\rho_{\Lambda} \sim 2.5 \times 10^{-11}~[\mathrm{eV^{4}}]$.
In using other gas species, one can multiply the ratio of molecular weight to that of hydrogen molecule with the above $(- \frac{T^{\mu}_{\mu}}{\rho_{\Lambda}})$ value.
As a benchmark, in Eq.~\eqref{Eq: chameleon mass formula} we may 
choose $\alpha = 10^{4}~[\mathrm{eV^{-2}}]$ as in Fig.~\ref{Fig: chameleon mass plot}.
For the range $\lambda=2$--$200$, 
the case $n=1$ predicts the chameleon in the testable range as in Fig.~\ref{Fig: sensitivity}.
The case $n=1/2$ is also testable by raising the gas pressure or changing the gas species,
although larger $\lambda$ for $n=1/2$ is testable even in the current setup.
Note that the case $n=2$ predicts heavy masses outside the sensitivity $m_{\varphi} \simeq 4.1 \times 10^{-3}~[\mathrm{eV}]$.

\section{Feasibility to extract the chameleon character}
\begin{figure}[htbp]
\centering
\includegraphics[width=0.4\textwidth]{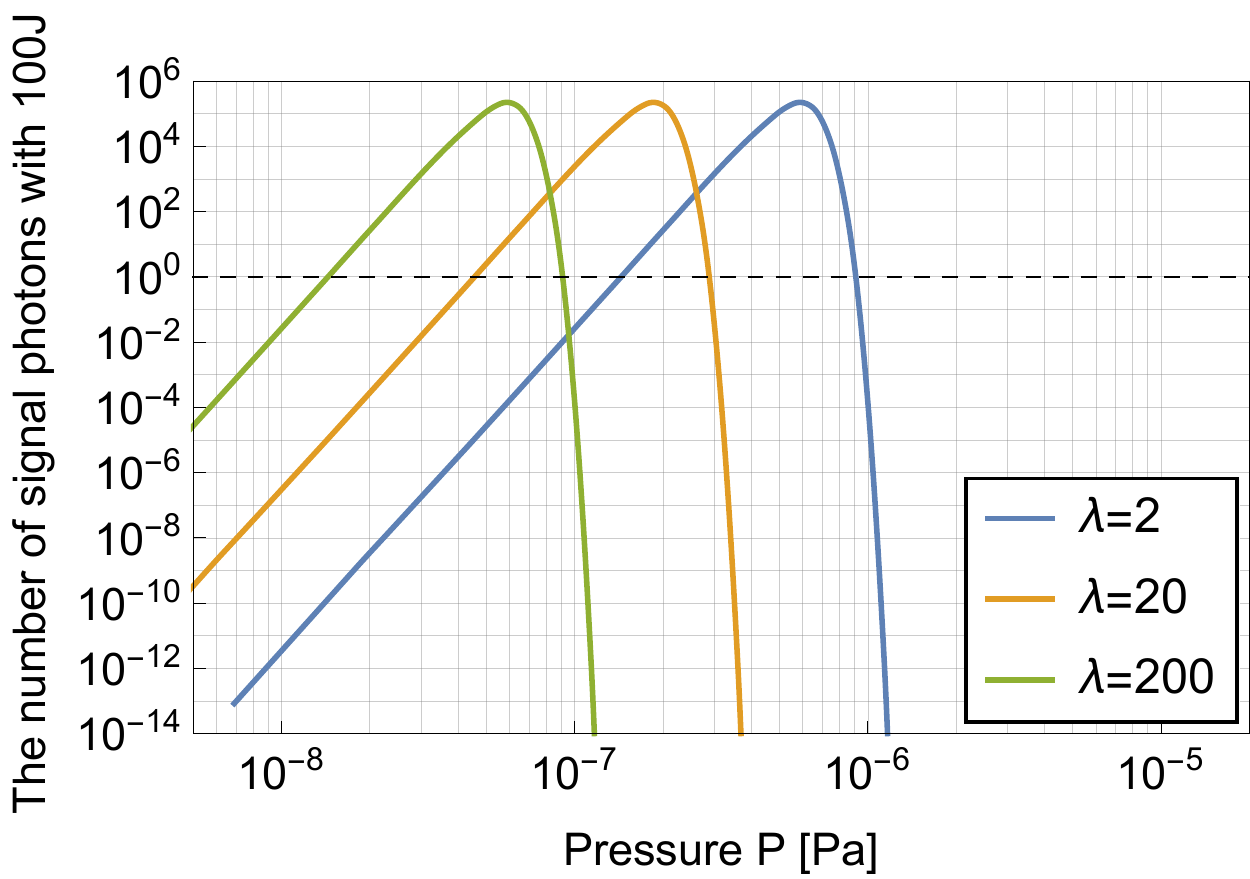}
\caption{
Expected number of stimulated signal photons as a function of gas pressure comprising hydrogen molecules at $300~[\mathrm{K}]$ for an $n=1$ with $\lambda =$ 2, 20, and 200 cases. 
The signal yield is calculated by assuming the photon-chameleon coupling defined in Eq.~\eqref{Eq: photon coupling} 
and the set of experimental parameters used in~\cite{Homma:2019rqb}, with the same pulse energy of $100~[\mathrm{J}]$ for the two beams.
The dashed horizontal line shows unity above which experiments can test the characteristic features of chameleon models. 
To evaluate the dependence for other types of molecules with molar mass $M$, 
one can scale the pressure values by multiplying $M_{{H}_{2}}/M$ with the molar mass of hydrogen molecule $M_{{H}_{2}}$.}
\label{Fig: yield}
\end{figure}

The key feature of the chameleon exchange appears in its density dependence of the stimulated resonant scattering rate. 
Figure~\ref{Fig: yield} shows the expected number of signal photons as a function of gas pressure comprising hydrogen molecules at $300~[\mathrm{K}]$ for an $n=1$ with $\lambda =$ 2, 20, and 200 cases. 
The signal yield is proportional to $\tilde{\beta_{\gamma}}^2$ with the set of parameters used in~\cite{Homma:2019rqb},
which assumed two beams with the same pulse energy of $100~[\mathrm{J}]$.~\footnote{
Because we know the mass-coupling relation from Ref.~\cite{Homma:2019rqb} in advance, 
we can evaluate the coupling strength as a function of pressure through Eq.~\eqref{Eq: chameleon mass formula},
and then eventually scale the yield $N_{obs}=100$ with the coupling strength as a function of pressure. 
}
The pressure range where the signal yield exceeds unity is the region of interest to test chameleon models.

As for background processes, four-wave mixing (FWM) via residual atoms produces photon energies that are kinematically similar to those of the signal via energy-momentum conservation~\cite{Druet:FWM}. 
The photon yield from the atomic FWM  process is known to be proportional to the square of the third order polarization susceptibility $\chi^{(3)}$. 
Because $\chi^{(3)}$ is proportional to the number density of the atoms, and hence to gas pressure, 
the atomic FWM yield is expected to have a quadratic pressure dependence, 
and this dependence has indeed been observed in the actual searching setup in laser frequencies~\cite{Hasebe:2015jxa,Nobuhiro:2020fub,Homma:2021hnl}, 
although the yield is expected to be negligibly small.~\footnote{
The FWM yield is proportional to the cubic of beam peak intensity $[\mathrm{J/s/m^2}]$ at a focal point. 
Based on the expected yield in the most recent search with pulsed lasers with several $10$'s femtosecond duration 
and photon energies of $\sim 1~[\mathrm{eV}]$~\cite{Homma:2021hnl}, 
and with the intensity scaling to much lower peak power beams with a few nanosecond duration and the lower photon energies of $\sim 10^{-5}~[\mathrm{eV}]$ assumed in~\cite{Homma:2019rqb}, 
a negligible amount of FWM photons from the atomic process is predicted if $\chi^{(3)}$ values in the two microwave frequencies are similar to those in the laser frequencies.
}

We further note that only the possible standard model process in vacuum, light-by-light scattering induced by a box diagram in quantum electrodynamics, is also negligible in the mass range $(0.1-1)~[\mu \mathrm{eV}]$ due to the sixth-power dependence of the cross section on the center-of-mass system energy even if we take the effect of stimulation into account~\cite{Homma:2017cpa}. 
The power exponent on the lower pressure side is around 10 for $n=1$, which is clearly distinguishable from the quadratic pressure dependence of the background atomic FWM process and also separable from pressure-independent BSM signals, if there are any. 
Therefore, in principle, the proposed method can provide firm grounds to purely test the chameleon mechanism separably from backgrounds in laboratory experiments.

\section{Conclusion}
In conclusion, hunting chameleons as DE is possible by measuring the pressure-dependent stimulated photon-photon scattering in a laboratory-based radar-collider experiment. 
If two focused radar pulses with $100~[\mathrm{J/ns}]$ are each available as designed in~\cite{Homma:2019rqb}, 
the chameleon model with a cosmologically allowed parameter space, $n=1$ with a wide range of $\lambda=2$--$200$, 
is testable based on the tenth-power steep pressure dependence of the signal yield with a chameleon mass in the range $(0.1-1)~[\mu \mathrm{eV}]$. 
This observable can clearly discriminate chameleons from other background processes. 
Therefore, the proposed method can provide a unique opportunity to strictly constrain the viable $F(R)$ gravity models for DE,
which can be consistent with the current cosmological observations. 
Given future developments on the microwave technologies and a capability for accurate pressure control, 
this method will pave the way to directly unveil dark energy in the completely controlled manner.

\begin{acknowledgments}
This work is dedicated to the late professor Yasunori Fujii, whose works on a dynamical dark energy model motivated us to start our collaboration. 
T.K. is grateful to Hitoshi Katsuragawa for the fruitful discussion and valuable advice and is supported by the National Key R\&D Program of China (2021YFA0718500).
S.M. work was supported in part by the National Science Foundation of China (NSFC) under Grants No.~11747308, No.~11975108, and No.~12047569 
and the Seeds Funding of Jilin University. 
K.H. acknowledges the support of the Collaborative Research Program of the Institute for Chemical Research of Kyoto University (Grant No.~2021-88) 
and Grants-in-Aid for Scientific Research No.~19K21880 and No.~21H04474 from the Ministry of Education, Culture, Sports, Science and Technology (MEXT) of Japan.
\end{acknowledgments}  

\appendix

\section{Electromagnetic Field in a Focused Photon Beam}
\subsection{Setup for Focusing Optics}
We consider the electric field $\mathbf{E}$ and magnetic field $\mathbf{B}$ in the focusing optics.
The wave equation in vacuum goes like 
\begin{align}
\Box \mathbf{F}(\mathbf{r},t) =0
\label{Eq: wave equation}
\, .
\end{align}
Its solution $\mathbf{F}(\mathbf{r}, t)$ is expressed in terms of the complex amplitude as 
\begin{align}
\mathbf{F}(\mathbf{r}, t) = \frac{1}{2} \hat{\mathbf{F}}(\mathbf{r}) e^{- i \omega t} + {\rm c.c}.
\label{Eq: wave equation sol}
\, .
\end{align}
Then the amplitude $\hat{\mathbf{F}}(\mathbf{r})$ satisfies the Helmholtz equation, 
\begin{align}
\left( \nabla^{2} + k^{2} \right) \hat{\mathbf{F}} (\mathbf{r}) = 0
\label{Eq: Helmholtz equation}
\, ,
\end{align}
where $k=\omega/c$ is the wave number $k=|\mathbf{k}|$ with 
$\mathbf{k} = (k_{x}, k_{y}, k_{z}) $.  
Because $k_{z} = \sqrt{k^2 - k^{2}_{x} - k^{2}_{y}}$ is determined for a given wave number, the Fourier expansion of $\hat{\mathbf{F}}(\mathbf{r})$ goes like 
\begin{align}
\hat{\mathbf{F}} (\mathbf{r}) = \int^{\infty}_{-\infty} \int^{\infty}_{-\infty} dk_{x} dk_{y} \mathbf{f}(k_{x}, k_{y}, z) e^{i\left(k_{x}x + k_{y}y \right)}
\label{Eq: Fourier expansion}
\, .
\end{align}
By substituting Eq.~\eqref{Eq: Fourier expansion} into Eq.~\eqref{Eq: Helmholtz equation},
the Fourier kernel function ${\bf f}(k_{x}, k_{y}, z)$ is written as
\begin{align}
\mathbf{f}(k_{x}, k_{y}, z) = \mathbf{U}(k_{x}, k_{y}) e^{ik_{z}z} + \mathbf{V}(k_{x}, k_{y}) e^{-ik_{z}z}
\label{Eq: Helmholtz equation sol}
\, ,
\end{align}
so that 
\begin{align}
\begin{split}
\hat{\mathbf{F}} (\mathbf{r}) =& 
\int^{\infty}_{-\infty} \int^{\infty}_{-\infty} dk_{x} dk_{y} \left[  \mathbf{U}(k_{x}, k_{y}) e^{ik_{z}z} \right.
\\
& \qquad \left.
+ \mathbf{V}(k_{x}, k_{y}) e^{-ik_{z}z} \right] e^{i\left(k_{x}x + k_{y}y \right)}
\label{Eq: amplitude}
\, .
\end{split}
\end{align}

\begin{figure}[htbp]
\centering
\includegraphics[width=0.4\textwidth]{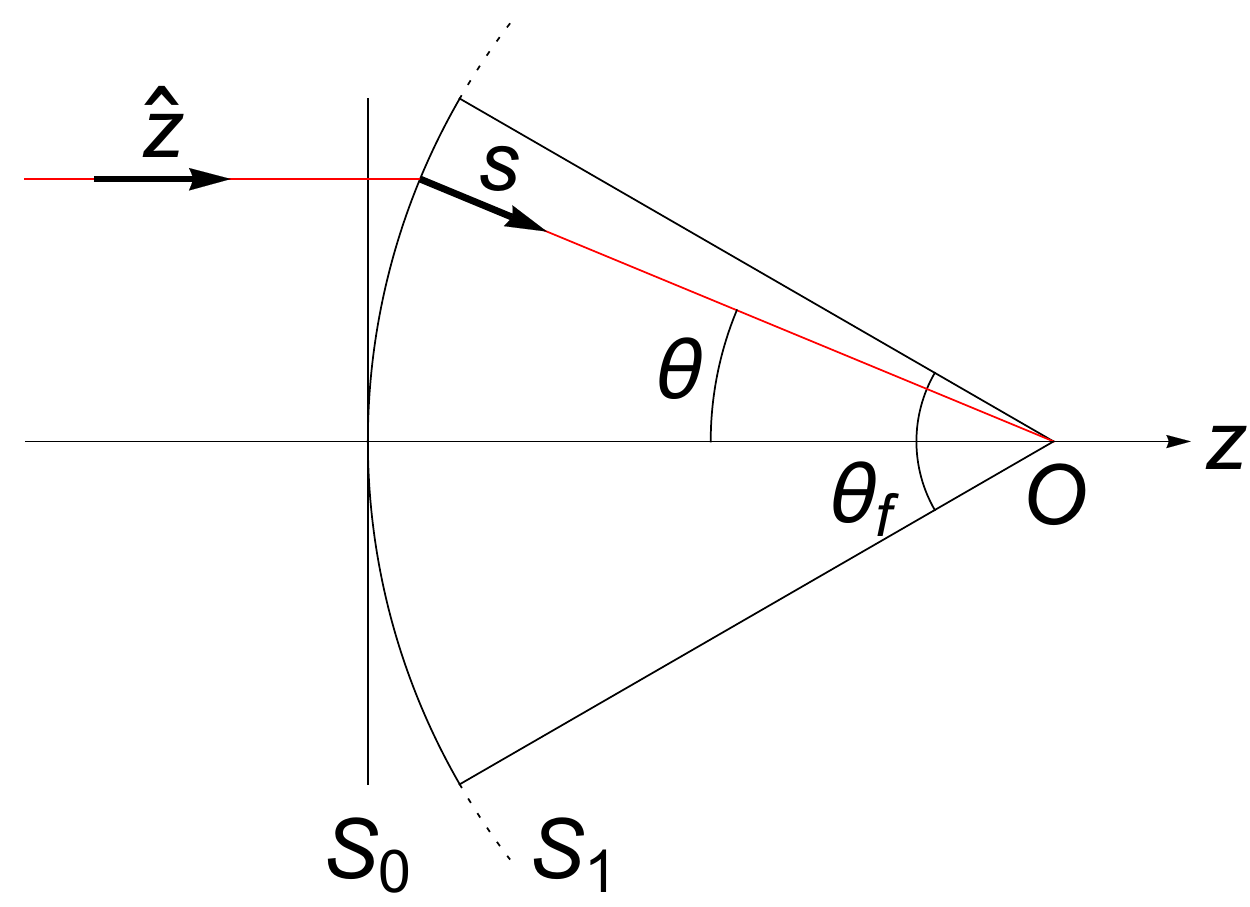}
\caption{
A schematic picture of focusing optics. The red line 
follows the photon beam.  
The incident plane wave along $z$-axis entering the plane $S_{0}$ is converted into the spherical wave with the unit vector $\mathbf{s}$ perpendicular to the sphere $S_{1}$, which is measured by the angle $\theta$ from the $z$-axis, 
and focused on the origin $O$.  
Here $\theta_{f}$ is the aperture angle.
}
\label{Fig: setup}
\end{figure}

Next, we consider the case of focusing the photon beam.
See Fig.~\ref{Fig: setup}. 
Writing the electromagnetic field of the focused photon beam with the frequency $\omega$ as
\begin{align}
\begin{split}
\mathbf{E}(\mathbf{r}, t) &= \frac{1}{2} \hat{\mathbf{E}}(\mathbf{r}) e^{-\omega t} + c.c.
\\
\mathbf{B}(\mathbf{r}, t) &= \frac{1}{2} \hat{\mathbf{B}}(\mathbf{r}) e^{-\omega t} + c.c. 
\, ,
\end{split}
\end{align}
we pay attention to the complex amplitudes $\hat{\mathbf{E}}(\mathbf{r})$ and  $\hat{\mathbf{B}}(\mathbf{r})$.
Eq.~\eqref{Eq: amplitude} allows us to express the amplitude of the electric field $\hat{\mathbf{E}}(\mathbf{r})$ as 
\begin{align}
\begin{split}
\hat{\mathbf{E}} (\mathbf{r}) & =
\int^{\infty}_{-\infty} \int^{\infty}_{-\infty} ds_{x} ds_{y} \left[  \mathbf{U}(s_{x}, s_{y}) e^{ik s_{z}z} \right.
\\
& \qquad \left.
+ \mathbf{V}(s_{x}, s_{y}) e^{-ik s_{z}z} \right] e^{ik\left(s_{x}x + s_{y}y \right)}
\, ,
\end{split}
\end{align}
where $s_{z} = \sqrt{1 - s^{2}_{x} - s^{2}_{y}}$.
Considering the case where $s^{2}_{x} + s^{2}_{y}\leq1$, we find 
\begin{align}
\begin{split}
\hat{\mathbf{E}} (\mathbf{r}) & =
\iint_{s^{2}_{x} + s^{2}_{y}\leq1} ds_{x} ds_{y} \left[  \mathbf{U}(s_{x}, s_{y}) e^{ik s_{z}z} \right.
\\
& \qquad \left.
+ \mathbf{V}(s_{x}, s_{y}) e^{-ik s_{z}z} \right] e^{ik\left(s_{x}x + s_{y}y \right)}
\label{Eq: electric field}
\, .
\end{split}
\end{align}
By introducing the variables,
\begin{align}
u_{x} = \frac{x}{f}\, , \ 
u_{y} = \frac{y}{f}\, , \ 
u_{z} = \frac{z}{f} = -\sqrt{1-u^2_{x} - u^2_{y}}
\, ,
\end{align}
$\hat{\mathbf{E}} (\mathbf{r})$ at a point $\mathbf{r}_{1} = (x_{1},y_{1},z_{1})$ on the sphere $S_{1}$ is evaluated as 
\begin{align}
\begin{split}
\hat{\mathbf{E}} (\mathbf{r}_{1}) & =
\iint_{s^{2}_{x} + s^{2}_{y}\leq1} ds_{x} ds_{y} \left[ \mathbf{U}(s_{x}, s_{y}) e^{ik f \Phi_{+} (s_{x}, s_{y})}  \right.
\\
& \qquad \left.
+ \mathbf{V}(s_{x}, s_{y}) e^{-ik f \Phi_{-}(s_{x}, s_{y})} \right] 
\label{Eq: electric field1}
\, ,
\end{split}
\end{align}
where $f$ stands for the focal distance, and
\begin{align}
\Phi_{\pm} (s_{x}, s_{y}) = s_{x}x + s_{y}y \pm \sqrt{1-s^2_{x} - s^2_{y}} u_{z}
\, .
\end{align}

Assuming the focal distance is larger than the wave length $fk\gg1$,
we apply the stationary phase method in evaluating the integration in Eq.~\eqref{Eq: electric field1}.
We first consider the first term in Eq.~\eqref{Eq: electric field1}, to find that a stationary point $(s_{x_{0}}, s_{y_{0}})$ satisfies 
\begin{align}
\begin{split}
\frac{\partial \Phi_{+} (s_{x}, s_{y}) }{\partial s_{x}} &= u_{x} - \frac{s_{x}}{\sqrt{1-s^2_{x} - s^2_{y}}} u_{z} = 0
\, , \\
\frac{\partial \Phi_{+} (s_{x}, s_{y}) }{\partial s_{y}} &= u_{y} - \frac{s_{y}}{\sqrt{1-s^2_{x} - s^2_{y}}} u_{z} = 0
\, .
\end{split}
\end{align}
Because $s_{x_{0}}<0$, one finds
\begin{align}
s_{x_{0}} = - u_{x}
\, , \ 
s_{y_{0}} = - u_{y}
\, .
\end{align}
The Taylor expansion of $\Phi_{+} (s_{x}, s_{y})$ around the stationary point $(s_{x_{0}}, s_{y_{0}})$ gives
\begin{align}
\begin{split}
&\Phi_{+} (s_{x}, s_{y}) 
\\
&= \Phi_{+} (s_{x_{0}}, s_{y_{0}})
+ \frac{w_{1}}{2} (s_{x} - s_{x_{0}})^{2}
\\
& \quad
+ \frac{w_{2}}{2} (s_{y} - s_{y_{0}})^{2}
+ w_{3} (s_{x} - s_{x_{0}})(s_{y} - s_{y_{0}})
+ \cdots
\, ,
\end{split}
\end{align}
where we have defined 
\begin{align}
\begin{split}
w_{1} 
&= - \frac{1-u^{2}_{y}} {\left(1-s^2_{x} - s^2_{y} \right)^{3/2}} u_{z} 
\, , \\
w_{2} 
&= - \frac{1-u^{2}_{x}} {\left(1-s^2_{x} - s^2_{y} \right)^{3/2}} u_{z} 
\, , \\
w_{3} 
&=  - \frac{u_{x}u_{y}} {\left(1-s^2_{x} - s^2_{y} \right)^{3/2}} u_{z} 
\, ,
\end{split}
\end{align}
with $u_{z}<0$, $w_{1}>0$, and $w_{2}>0$. 

When $kf \gg 1$, the integrand in the vicinity of the stationary point $(s_{x_{0}}, s_{y_{0}})$ mainly contributes to the integration,
and thus, we find
\begin{align}
\begin{split}
&\mathbf{W}_{1} (\mathbf{r}_{1})
\\
& \equiv \iint_{s^{2}_{x} + s^{2}_{y}\leq1} ds_{x} ds_{y} \mathbf{U} (s_{x}, s_{y}) e^{ik f \Phi_{+} (s_{x}, s_{y})} 
\\
& \approx  
\mathbf{U} (s_{x_{0}}, s_{y_{0}}) e^{ik f \Phi_{+} (s_{x_{0}}, s_{y_{0}})} 
\iint_{s^{2}_{x} + s^{2}_{y}\leq1} ds_{x} ds_{y} 
\\
& \quad
e^{ik f \left[  \frac{w_{1}}{2} (s_{x} - s_{x_{0}})^{2} + \frac{w_{2}}{2} (s_{y} - s_{y_{0}})^{2} + w_{3} (s_{x} - s_{x_{0}})(s_{y} - s_{y_{0}}) \right]} 
\, .
\end{split}
\end{align}
The infinitesimal region $(s_{x} - s_{x_{0}})^{2} + (s_{y} - s_{y_{0}})^{2} \leq \epsilon \ll 1$ dominates over 
the target integration domain $s^{2}_{x} + s^{2}_{y}\leq1$. 
Defining $\epsilon_{x} \equiv s_{x} - s_{x_{0}}$ and $\epsilon_{y} \equiv s_{y} - s_{y_{0}}$,
we find
\begin{align}
\begin{split}
&\mathbf{W}_{1} (\mathbf{r}_{1}) 
\\
& =
\mathbf{U} (s_{x_{0}}, s_{y_{0}}) e^{ik f \Phi_{+} (s_{x_{0}}, s_{y_{0}})} 
\\
& \qquad \times  
\iint_{\epsilon^{2}_{x} + \epsilon^{2}_{y}\leq\epsilon} d\epsilon_{x} d\epsilon_{y} 
e^{ik f \left[  \frac{w_{1}}{2} \epsilon^{2}_{x} + \frac{w_{2}}{2} \epsilon^{2}_{y} + w_{3} \epsilon_{x} \epsilon_{y}  \right]} 
\\
& \approx 
\mathbf{U} (s_{x_{0}}, s_{y_{0}}) e^{ik f \Phi_{+} (s_{x_{0}}, s_{y_{0}})} 
\\
& \quad \times
\int^{\epsilon}_{-\epsilon} d\epsilon_{y} \int^{\sqrt{\epsilon^{2} -  \epsilon^{2}_{y}}}_{-\sqrt{\epsilon^{2} -  \epsilon^{2}_{y}} } d\epsilon_{x} 
e^{ik f \left[  \frac{w_{1}}{2} \epsilon^{2}_{x} + \frac{w_{2}}{2} \epsilon^{2}_{y} + w_{3} \epsilon_{x} \epsilon_{y}  \right]} 
\, .
\end{split}
\end{align}
Performing change of variables,  $X \equiv \sqrt{kf} \epsilon_{x}$ and  
$Y \equiv \sqrt{kf} \epsilon_{y}$, the integration can be approximated as
\begin{align}
\begin{split}
&\mathbf{W}_{1} (\mathbf{r}_{1}) 
\\
& \approx 
\mathbf{U} (s_{x_{0}}, s_{y_{0}}) e^{ik f \Phi_{+} (s_{x_{0}}, s_{y_{0}})} 
\\
& \quad \times
\frac{1}{kf}
\int^{\infty}_{-\infty} dY \int^{\infty}_{-\infty} dX 
e^{i \left[  \frac{w_{1}}{2} X^{2} + \frac{w_{2}}{2} Y^{2} + w_{3} XY  \right]} 
\, .
\end{split}
\end{align}
Noting 
\begin{align}
\int^{\infty}_{-\infty} dY \int^{\infty}_{-\infty} dX 
e^{i \left[  \frac{w_{1}}{2} X^{2} + \frac{w_{2}}{2} Y^{2} + w_{3} XY  \right]} 
= -2\pi i u_{z}
\, ,
\end{align}
and using $\Phi_{+} (s_{x_{0}}, s_{y_{0}}) = -1$, we find
\begin{align}
\mathbf{W}_{1} (\mathbf{r}_{1}) 
& \approx - \frac{2\pi i u_{z}}{kf} U(-u_{x}, -u_{y}) e^{- ikf} 
\,. 
\end{align}

In the same way, we can compute the second term in Eq.~\eqref{Eq: electric field1}. 
Thus we find 
\begin{align}
\hat{\mathbf{E}} (\mathbf{r}_{1}) \approx
\frac{2\pi i u_{z}}{kf} \left[ - \mathbf{U}(-u_{x}, -u_{y}) e^{- ikf} + \mathbf{V}(u_{x}, u_{y}) e^{ikf} \right]
\,. 
\end{align}
If there is no aberration in the focusing optics so that 
\begin{align}
\begin{split}
s_{x} &= - \frac{x_{1}}{f} = -u_{x}
\, , \\
s_{x} &= - \frac{y_{1}}{f} = -u_{y}
\, , \\
s_{x} &= - \frac{z_{1}}{f} = -u_{z}
\, , 
\end{split}
\end{align}
then we have 
\begin{align}
\hat{\mathbf{E}} (\mathbf{r}_{1}) \approx
\frac{2\pi i s_{z}}{kf} \left[  - \mathbf{U}(s_{x}, s_{y}) e^{- ikf} + \mathbf{V}(-s_{x}, -s_{y}) e^{ikf} \right] 
\,. 
\label{Eq: electric field2}
\end{align}

Next, we compute the coefficients $U(s_{x}, s_{y})$ and $ V(-s_{x}, -s_{y})$ in Eq.~\eqref{Eq: electric field2}.
Because the amplitude of the spherical wave is proportional to the inverse of distance, 
we can express $\hat{E} (\mathbf{r}_{1}) $ as
\begin{align}
\hat{\mathbf{E}} (\mathbf{r}_{1}) = \frac{{\bf  a}_{1}(s_{x}, s_{y})}{f} e^{ik\phi(\mathbf{r}_{1})}
\, .
\end{align}
Here, $\mathbf{a}_{1}(s_{x}, s_{y})$ is a vector perpendicular to the light ray from the sphere $S_{1}$ to the focus $O$,
and $\phi(\mathbf{r}_{1})$ is the eikonal function to describe the optical path length from a point $O^{\prime}$ to $\mathbf{r}_{1}$ in the object space.
Assuming the optical path length of the spherical wave with its source at point $O^{\prime}$ 
which converges at the focus $O$ through the point $\mathbf{r}_{1}$ is expressed by the constant $C$ to be 
\begin{align}
\phi(\mathbf{r}_{1}) = C - |\mathbf{r}_{1}| = C - f
\, .
\end{align}
Thus, we find the relation between ${\bf a}_1(s_{x}, s_{y})$ and the coefficients $\mathbf{U}(s_{x}, s_{y})$, $\mathbf{V}(-s_{x}, -s_{y})$ as follows
\begin{align}
\begin{split}
\mathbf{U}(s_{x}, s_{y}) &= - \frac{ik}{2\pi} \frac{\mathbf{a}_{1}(s_{x}, s_{y})}{s_{z}} e^{ikC}\
\\
\mathbf{V}(s_{x}, s_{y}) &= 0
\, .
\end{split}
\end{align}
Ignoring the constant factor $e^{ikC}$ and substituting the above into Eq.~\eqref{Eq: electric field}, 
we obtain the amplitude of the electric field,
\begin{align}
\hat{\mathbf{E}} (\mathbf{r}) =
- \frac{ik}{2\pi} \iint_{s^{2}_{x} + s^{2}_{y}\leq1} ds_{x} ds_{y} \frac{\mathbf{a}_{1}(s_{x}, s_{y})}{s_{z}} e^{ik \mathbf{s} \cdot \mathbf{r}}
\label{Eq: electric field def}
\, ,
\end{align}
and, in a similar way, that of the magnetic field goes like  
\begin{align}
\hat{\mathbf{B}} (\mathbf{r}) =
- \frac{ik}{2\pi} \iint_{s^{2}_{x} + s^{2}_{y}\leq1} ds_{x} ds_{y} \frac{\mathbf{a}_{2}(s_{x}, s_{y})}{s_{z}} e^{ik \mathbf{s} \cdot \mathbf{r}}
\label{Eq: magnetic field def}
\, .
\end{align}

Finally, we rewrite ${\bf a}_{1}(s_{x}, s_{y})$ and ${\bf a}_{2}(s_{x}, s_{y})$ in terms of angles in the polar coordinate.
Considering $\mathbf{e}_{0}(\theta, \varphi)$ as the electric field (with the phase removed) of the incident light ray on the plane $S_{0}$
and $\mathbf{e}_{1}(\theta, \varphi)$ as that on the sphere $S_{1}$,
we write $\hat{\mathbf{e}}_{0}(\varphi)$ and $\hat{\mathbf{e}}_{1}(\theta, \varphi)$ 
as unit vectors parallel to $\mathbf{e}_{0}(\theta, \varphi)$ and $\mathbf{e}_{1}(\theta, \varphi)$, respectively.  
One then finds
\begin{align}
\mathbf{e}_{1}(\theta, \varphi) = \frac{\mathbf{a}_{1} (\theta, \varphi)}{f}
\label{Eq: electric field basis}
\, .
\end{align}
Let $\delta S_{0}$ be the infinitesimal circular ring on $S_{0}$, and $\delta S_{1}$ be the area of the projection of $\delta S_{0}$ onto $S_{1}$. 
Then the following relation holds between $\delta S_{0}$ and $\delta S_{1}$: 
\begin{align}
\delta S_{0} = \delta S_{1} \cos \theta
\, .
\end{align}
Using the energy conservation law
\begin{align}
|\mathbf{e}_{0}(\theta, \varphi)|^{2} \delta S_{0} = |\mathbf{e}_{1}(\theta, \varphi)|^{2} \delta S_{1}
\, ,
\end{align}
we obtain
\begin{align}
\mathbf{e}_{1}(\theta, \varphi) = |\mathbf{e}_{0}(\theta, \varphi)| \cos^{\frac{1}{2}} \theta \, \hat{\mathbf{e}}_{1}(\theta, \varphi) 
\,. 
\end{align}

\begin{figure}[htbp]
\centering
\includegraphics[width=0.4\textwidth]{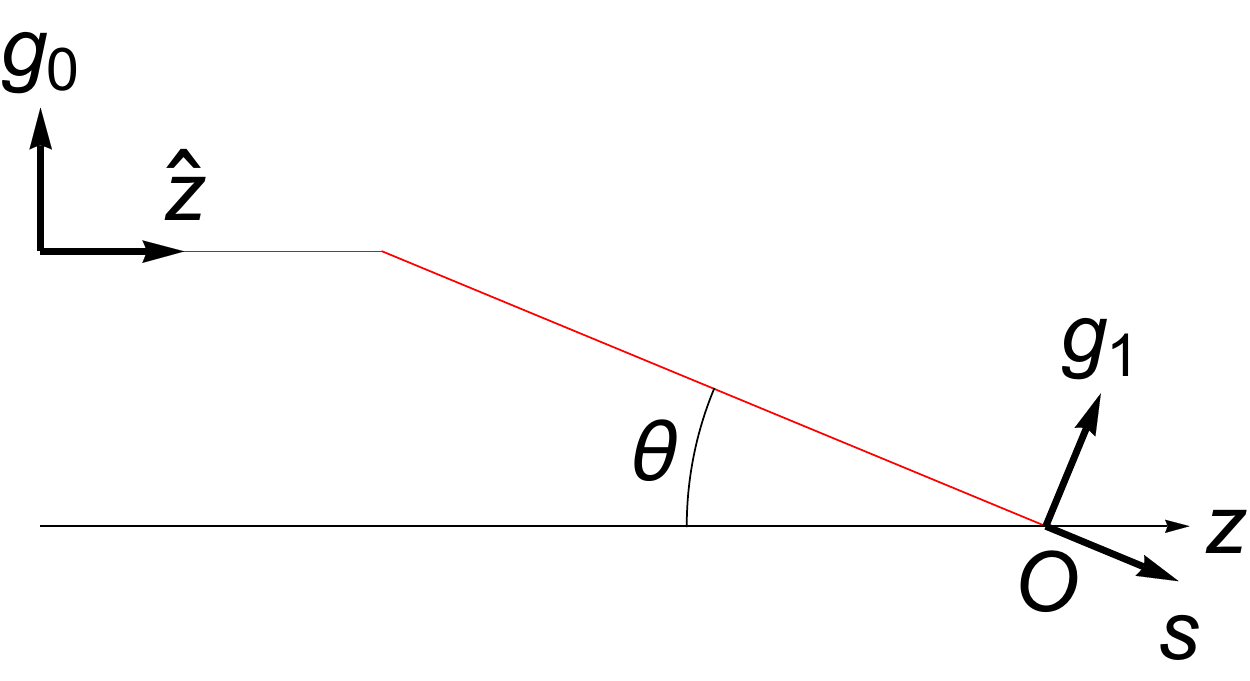}
\includegraphics[width=0.4\textwidth]{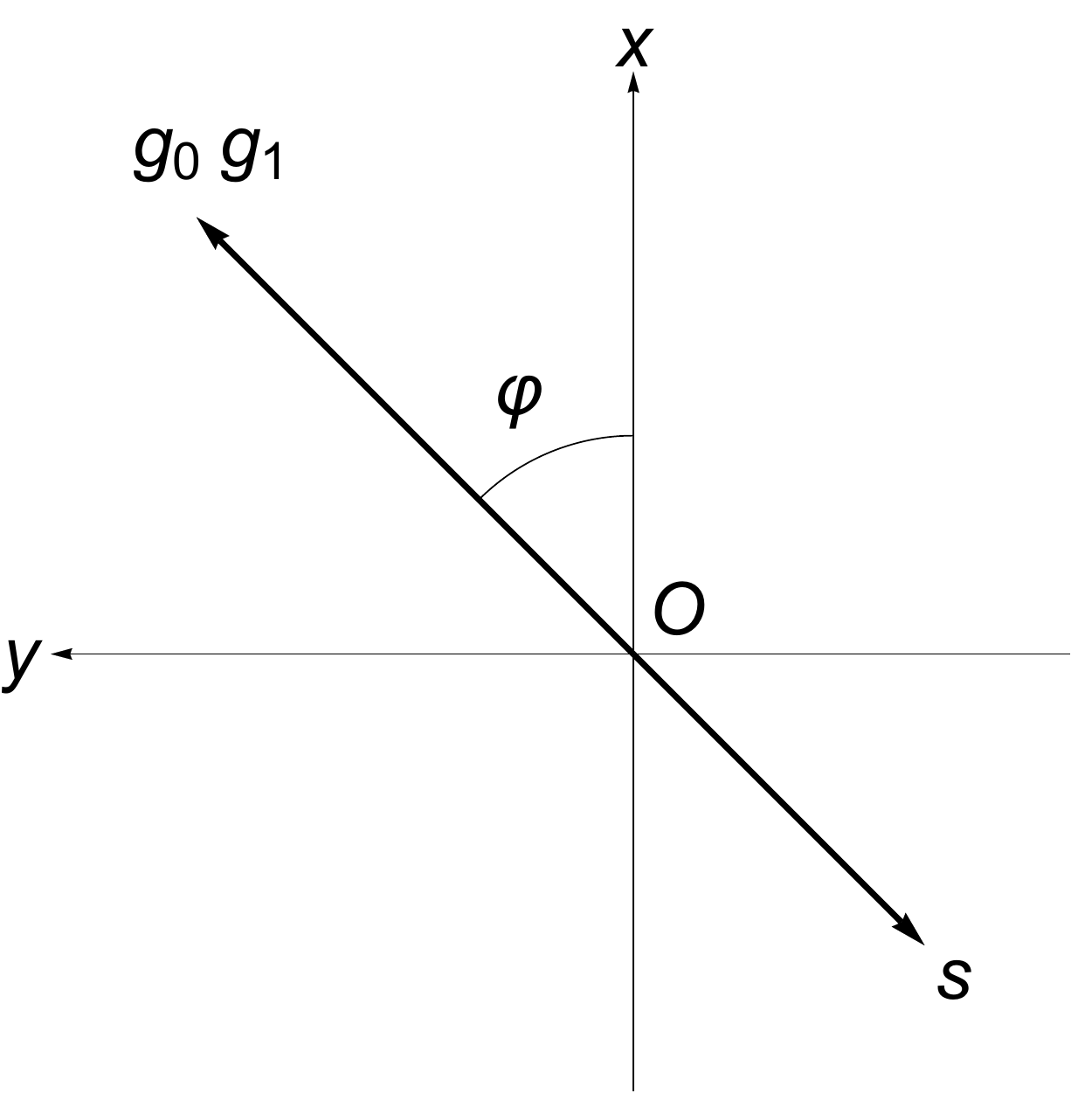}
\caption{
Alignment of vectors and coordinate system. The detailed description is provided in the text. 
}
\label{Fig: vectors}
\end{figure}

Let the two vectors $\mathbf{g}_{0}(\varphi)$ and $\mathbf{g}_{1}(\theta, \varphi)$ be unit vectors,  
that are perpendicular to the incident light ray and the focused light ray, respectively. 
They lie on a plane including both the light ray and the optical axis (see  Fig.~\ref{Fig: vectors}). 
$\mathbf{g}_{0}(\varphi)$ and $\mathbf{g}_{1}(\theta, \varphi)$ are given as
\begin{align}
\begin{split}
\mathbf{g}_{0}(\varphi) &= \cos \varphi \, \hat{\mathbf{x}} + \sin \varphi \, \hat{\mathbf{y}} 
\\
\mathbf{g}_{1}(\theta, \varphi) &= \cos \theta \cos \varphi \, \hat{\mathbf{x}} + \cos \theta \sin \varphi \, \hat{\mathbf{y}} + \sin \theta \, \hat{\mathbf{z}} 
\\ 
\mathbf{s}(\theta, \varphi) &= - \sin \theta \cos \varphi \, \hat{\mathbf{x}} - \sin \theta \sin \varphi \, \hat{\mathbf{y}} + \cos \theta \, \hat{\mathbf{z}} 
\label{Eq: basis transformation}
\, ,
\end{split}
\end{align}
where $\hat{\mathbf{x}}$, $\hat{\mathbf{y}}$, and $\hat{\mathbf{z}}$ are $x, y, z$-direction unit vectors.
By definition, we have 
\begin{align}
\begin{split}
\mathbf{g}_{0}(\varphi) \times \hat{\mathbf{z}} 
&=
\mathbf{g}_{1}(\theta, \varphi) \times \mathbf{s}(\theta, \varphi) 
\\
& =
\sin \varphi \, \hat{\mathbf{x}} - \cos \varphi  \, \hat{\mathbf{y}}
\, .
\end{split}
\end{align}
Because $\hat{\mathbf{e}}_{0}(\varphi)$ is perpendicular to the $z$-axis, 
we can decompose it as
\begin{align}
\begin{split}
\hat{\mathbf{e}}_{0}(\varphi) 
&= 
\left[ \hat{\mathbf{e}}_{0}(\varphi) \cdot \mathbf{g}_{0}(\varphi) \right] \mathbf{g}_{0}(\varphi) 
\\
& \qquad 
+ \left[ \hat{\mathbf{e}}_{0}(\varphi) \cdot \left(\mathbf{g}_{0}(\varphi) \times \hat{\mathbf{z}} \right) \right] \left( \mathbf{g}_{0}(\varphi) \times \hat{\mathbf{z}} \right)
\, .
\end{split}
\end{align}
When $\mathbf{e}_{0}(\theta, \varphi)$ is converted into $\mathbf{e}_{1}(\theta, \varphi)$,
the $\mathbf{g}_{0}(\varphi)$-direction component is transformed into the $\mathbf{g}_{1}(\varphi)$-direction component,
and the $\mathbf{g}_{0}(\varphi) \times \hat{\mathbf{z}}$-direction component is done into the $\mathbf{g}_{1}(\theta, \varphi) \times \mathbf{s}(\theta, \varphi)$-direction component. 
Thus, 
\begin{align}
\begin{split}
&\mathbf{e}_{1}(\theta, \varphi) 
\\
&= 
|\mathbf{e}_{0}(\theta, \varphi)| \cos^{\frac{1}{2}} \theta 
\left \{ \left[ \hat{\mathbf{e}}_{0}(\varphi) \cdot \mathbf{g}_{0}(\varphi) \right] \mathbf{g}_{1}(\varphi) \right.
\\
& \qquad  \left.
+ \left[ \hat{\mathbf{e}}_{0}(\varphi) \cdot (\mathbf{g}_{0}(\varphi) \times \hat{\mathbf{z}}) \right] 
\left( \mathbf{g}_{1}(\theta, \varphi) \times \mathbf{s}(\theta, \varphi)  \right) \right\}
\, .
\end{split}
\end{align}
In the focusing optics without aberration, 
we can express $\mathbf{e}_{0}(\theta, \varphi)$ as
\begin{align}
\mathbf{e}_{0}(\theta, \varphi) = e_{0} l_{0}(\theta) \hat{\mathbf{e}}_{0}(\varphi) 
\label{Eq: basis coordinate decomposition}
\, ,
\end{align}
where $e_{0}$ is the maximum value of $|\mathbf{e}_{0}(\theta, \varphi)|$, and $l_{0}(\theta)$ is the relative amplitude.
Therefore, by using Eq.~\eqref{Eq: electric field basis}, we obtain
\begin{align}
\begin{split}
& \mathbf{a}_{1} (\theta, \varphi)
\\
&= 
f e_{0} l_{0}(\theta) \cos^{\frac{1}{2}} \theta 
\left \{ \left[ \hat{\mathbf{e}}_{0}(\varphi) \cdot \mathbf{g}_{0}(\varphi) \right] \mathbf{g}_{1}(\theta, \varphi) \right.
\\
& \qquad \left.
+ \left[ \hat{\mathbf{e}}_{0}(\varphi) \cdot (\mathbf{g}_{0}(\varphi) \times \hat{\mathbf{z}}) \right] 
\left( \mathbf{g}_{1}(\theta, \varphi) \times \mathbf{s}(\theta, \varphi)  \right) \right\}
\, .
\end{split}
\end{align}
The relation between the electric and magnetic fields leads to
\begin{align}
\begin{split}
&\mathbf{a}_{2} (\theta, \varphi) 
\\
&= 
\mathbf{s}(\theta, \varphi) \times \mathbf{a}_{1} (\theta, \varphi) 
\\
&= 
f e_{0} l_{0}(\theta) \cos^{\frac{1}{2}} \theta 
\left \{ \left[ \hat{\mathbf{e}}_{0}(\varphi) \cdot (\mathbf{g}_{0}(\varphi) \times \hat{\mathbf{z}}) \right] \mathbf{g}_{1}(\theta, \varphi)  \right.
\\
& \qquad \left.
- \left[ \hat{\mathbf{e}}_{0}(\varphi) \cdot \mathbf{g}_{0}(\varphi) \right] \left( \mathbf{g}_{1}(\theta, \varphi) \times \mathbf{s}(\theta, \varphi) \right)
\right\}
\, .
\end{split}
\end{align}

In the cylindrical coordinate $(\rho, \varphi, z)$, $\mathbf{r}=\rho \cos \varphi \, \hat{\mathbf{x}} + \rho \sin \varphi  \, \hat{\mathbf{y}} + z \, \hat{\mathbf{z}}$,
and Eq.~\eqref{Eq: basis transformation} tells us
\begin{align}
\mathbf{s}(\theta^{\prime}, \varphi^{\prime}) \cdot \mathbf{r}
&= 
z \cos \theta^{\prime} 
- \rho \sin \theta^{\prime} \cos \left( \varphi^{\prime}  - \varphi  \right)
\, .
\end{align}
The Jacobian goes like 
\begin{align}
\frac{\partial\left( s_{x}, s_{y} \right)}{\partial\left( \theta^{\prime}, \varphi^{\prime} \right)} = \sin \theta^{\prime} \cos \theta^{\prime} 
\, , \ 
ds_{x} ds_{y} = s_{z} \sin \theta^{\prime} d\theta^{\prime} d\varphi^{\prime} 
\, . 
\end{align}
Thus Eqs.~\eqref{Eq: electric field def} and \eqref{Eq: magnetic field def} are written in the polar-coordinate as follows: 
\begin{align}
\begin{split}
\hat{\mathbf{E}} (\mathbf{r}) &=
- \frac{ik}{2\pi} \int^{\frac{\theta_{f}}{2}}_{0} d\theta^{\prime} \int^{2\pi}_{0} d\varphi^{\prime}  \mathbf{a}_{1}(\theta^{\prime}, \varphi^{\prime}) \sin \theta^{\prime}
\\
& \qquad \qquad \qquad \qquad
e^{ik \left[ z \cos \theta^{\prime} - \rho \sin \theta^{\prime} \cos \left( \varphi^{\prime}  - \varphi  \right) \right]}
\label{Eq: electric field def2}
\, ,
\end{split}
\\
\begin{split}
\hat{\mathbf{B}} (\mathbf{r}) &=
- \frac{ik}{2\pi} \int^{\frac{\theta_{f}}{2}}_{0} d\theta^{\prime} \int^{2\pi}_{0} d\varphi^{\prime}  \mathbf{a}_{2}(\theta^{\prime}, \varphi^{\prime}) \sin \theta^{\prime}
\\ 
& \qquad \qquad \qquad \qquad
e^{ik \left[ z \cos \theta^{\prime} - \rho \sin \theta^{\prime} \cos \left( \varphi^{\prime}  - \varphi  \right) \right]}
\label{Eq: magnetic field def2}
\, .
\end{split}
\end{align}

\subsection{Linear Polarization}
When the incident photon beam is linearly polarized,  
Eq.~\eqref{Eq: basis coordinate decomposition} goes like 
\begin{align}
\mathbf{e}_{0}(\theta, \varphi) = e_{0} l_{g}(\theta) \hat{\mathbf{x}}
\, , \ 
\hat{\mathbf{e}}_{0}(\varphi) = \hat{\mathbf{x}}
\, , 
\end{align}
where $l_{g}(\theta)$ expresses the Gaussian distribution,
\begin{align}
l_{g}(\theta) = \exp \left[ - \left(\frac{\sin \theta}{\sin \frac{\theta_{f}}{2}} \right)^{2} \right]
\, .
\end{align}
Using 
\begin{align}
\begin{split}
\hat{\mathbf{e}}_{0}(\varphi) \cdot \mathbf{g}_{0}(\varphi)  &= \cos \varphi  
\,, \\
\hat{\mathbf{e}}_{0}(\varphi) \cdot (\mathbf{g}_{0}(\varphi) \times \hat{\mathbf{z}}) &= \sin \varphi 
\,, 
\end{split}
\end{align}
we can derive each component of $\mathbf{a}_{1} (\theta, \varphi)$ and $\mathbf{a}_{2} (\theta, \varphi)$ in the Cartesian coordinate as follows:
\begin{align}
\begin{split}
&\mathbf{a}_{1} (\theta, \varphi)
\\
&= 
\frac{1}{2} f e_{0} l_{g}(\theta) \cos^{\frac{1}{2}} \theta 
\\
&  \qquad \times
\left \{ \left[  (1 + \cos \theta)  - (1 - \cos \theta) \cos 2 \varphi \right] \hat{\mathbf{x}} \right.
\\
& \qquad  \qquad \left.
+ (\cos \theta - 1) \sin 2 \varphi \, \hat{\mathbf{y}}
+ 2 \sin \theta \cos \varphi \, \hat{\mathbf{z}}  
\right\}
\, ,
\end{split}
\\
\begin{split}
&\mathbf{a}_{2} (\theta, \varphi) 
\\
&= 
\frac{1}{2} f e_{0} l_{g}(\theta) \cos^{\frac{1}{2}} \theta 
\\
&  \qquad \times
\left \{  (\cos \theta-1) \sin 2 \varphi  \, \hat{\mathbf{x}} \right.
\\
&  \qquad \qquad
+ \left[ \left( 1 + \cos \theta \right) + (1 - \cos \theta) \cos 2 \varphi  \right] \hat{\mathbf{y}}   
\\
&  \qquad \qquad \qquad \left.
+ 2 \sin \theta \sin \varphi  \, \hat{\mathbf{z}}
\right\}
\, .
\end{split}
\end{align}
Substituting the above into Eqs.~\eqref{Eq: electric field def2} and \eqref{Eq: magnetic field def2},
we finally obtain the amplitudes of the electric field and magnetic field as follows:
\begin{align}
\begin{split}
\hat{E}_{x} (\rho, \varphi, z) &= -i A \left[ I_{0} (\rho, z) + \cos 2\varphi I_{2} (\rho, z) \right]
\, , \\
\hat{E}_{y} (\rho, \varphi, z) &= -i A \sin 2\varphi I_{2} (\rho, z)
\, , \\
\hat{E}_{z} (\rho, \varphi, z) &= -2 A \cos \varphi I_{1} (\rho, z)
\, , \\
\hat{B}_{x} (\rho, \varphi, z) &= -i A \sin 2\varphi I_{2} (\rho, z)
\, , \\
\hat{B}_{y} (\rho, \varphi, z) &= -i A \left[ I_{0} (\rho, z) - \cos 2\varphi I_{2} (\rho, z) \right]
\, , \\
\hat{B}_{z} (\rho, \varphi, z) &= -2 A \cos \varphi I_{1} (\rho, z)
\label{Eq: linear elemag}
\, , 
\end{split}
\end{align}
where $A = kf e_{0}/2$, and 
\begin{align}
\begin{split}
I_{0} (\rho, z)
&=
\int^{\frac{\theta_{f}}{2}}_{0} d\theta^{\prime} l_{g}(\theta^{\prime}) \cos^{\frac{1}{2}} \theta^{\prime} \sin \theta^{\prime}
\\
& \qquad \qquad 
\times (1 + \cos \theta^{\prime}) J_{0}(k\rho \sin \theta^{\prime}) e^{ik z \cos \theta^{\prime} } 
\, ,
\\
I_{1} (\rho, z)
&=
\int^{\frac{\theta_{f}}{2}}_{0} d\theta^{\prime}  l_{g}(\theta^{\prime}) \cos^{\frac{1}{2}} \theta^{\prime}  \sin^{2} \theta^{\prime}   
\,, 
\\
& \qquad \qquad \qquad 
\times J_{1} (k\rho \sin \theta^{\prime}) e^{ik z \cos \theta^{\prime} }
\\
I_{2} (\rho, z)
&=
\int^{\frac{\theta_{f}}{2}}_{0} d\theta^{\prime} l_{g}(\theta^{\prime}) \cos^{\frac{1}{2}} \theta^{\prime} \sin \theta^{\prime}
\\
& \qquad \qquad 
\times (1 - \cos \theta^{\prime}) J_{2}(k\rho \sin \theta^{\prime})  e^{ik z \cos \theta^{\prime} } 
\label{Eq: linear integral func}
\, ,
\end{split}
\end{align}
where $J_{n}(x) \, (n=0, 1, 2)$ is Bessel function of the first kind.

\subsection{Circular Polarization}

When the incident photon beam is circularly polarized,
Eq.~\eqref{Eq: basis coordinate decomposition} goes like
\begin{align}
\begin{split}
\mathbf{e}_{0}(\theta, \varphi) &= e_{0} l_{g}(\theta) \left( \cos \varphi_{0} \hat{\mathbf{\rho}} - \sin \varphi_{0} \hat{\mathbf{\varphi}} \right)
\, , \\
\hat{\mathbf{e}}_{0}(\varphi) &= \left( \cos \varphi_{0} \hat{\bm{\rho}} - \sin \varphi_{0} \hat{\bm{\varphi}} \right)
\, , \
\varphi_{0} = \frac{\pi}{4}
\, , 
\end{split}
\end{align}
where $l_{bg}(\theta)$ expresses the Bessel-Gaussian distribution,
\begin{align}
l_{bg}(\theta) = J_{1} \left( \frac{\sin \theta}{\sin \frac{\theta_{f}}{2}} \right) \exp \left[ - \left(\frac{\sin \theta}{\sin \frac{\theta_{f}}{2}} \right)^{2} \right]
\, .
\end{align}
Here, one finds
\begin{align}
\hat{\bm{\rho}} = \cos \varphi \hat{\mathbf{x}} + \sin \varphi \hat{\mathbf{y}}
\, , 
\hat{\bm{\varphi}} = - \sin \varphi \hat{\mathbf{x}} + \cos \varphi \hat{\mathbf{y}}
\, .
\end{align}
In a way similar to the case of the linear polarization
we thus get 
\begin{align}
\begin{split}
\hat{E}_{x} (\rho, \varphi, z) 
&= -2 A \left[ \cos \varphi_{0} \cos \varphi U_{0} (\rho, z) \right.
\,, \\
& \left. \qquad \qquad 
+ \sin \varphi_{0} \sin \varphi U_{2} (\rho, z) \right]
\,, \\
\hat{E}_{y} (\rho, \varphi, z) 
&=  -2 A \left[ \cos \varphi_{0} \sin \varphi U_{0} (\rho, z) \right.
\,, \\
& \left. \qquad \qquad
- \sin \varphi_{0} \cos \varphi U_{2} (\rho, z) \right]
\,, \\
\hat{E}_{z} (\rho, \varphi, z) &= -2i A \cos \varphi_{0} U_{1} (\rho, z)
\,, \\
\hat{B}_{x} (\rho, \varphi, z) 
&=  -2 A \left[ \sin \varphi_{0} \cos \varphi U_{0} (\rho, z) \right.
\,, \\
& \left. \qquad \qquad
- \cos \varphi_{0} \sin \varphi U_{2} (\rho, z) \right]
\,, \\
\hat{B}_{y} (\rho, \varphi, z) 
&= -2 A \left[ \sin \varphi_{0} \sin \varphi U_{0} (\rho, z) \right.
\,, \\
& \left. \qquad \qquad 
+ \cos \varphi_{0} \cos \varphi U_{2} (\rho, z) \right]
\,, \\
\hat{B}_{z} (\rho, \varphi, z) &= -2i A \sin \varphi_{0} U_{1} (\rho, z)
\label{Eq: circular elemag}
\, ,
\end{split}
\end{align}
where 
\begin{align}
\begin{split}
U_{0} (\rho, z)
&=
\int^{\frac{\theta_{f}}{2}}_{0} d\theta^{\prime} l_{bg}(\theta^{\prime}) \cos^{\frac{3}{2}} \theta^{\prime} \sin \theta^{\prime}
\\
& \qquad \qquad \qquad 
J_{1}(k\rho \sin \theta^{\prime}) e^{ik z \cos \theta^{\prime} } 
\, , \\
U_{1} (\rho, z)
&=
\int^{\frac{\theta_{f}}{2}}_{0} d\theta^{\prime}  l_{bg}(\theta^{\prime}) \cos^{\frac{1}{2}} \theta^{\prime}  \sin^{2} \theta^{\prime}   
\\
& \qquad \qquad \qquad 
 J_{0} (k\rho \sin \theta^{\prime}) e^{ik z \cos \theta^{\prime} }
\, , \\
U_{2} (\rho, z)
&=
\int^{\frac{\theta_{f}}{2}}_{0} d\theta^{\prime} l_{bg}(\theta^{\prime}) \cos^{\frac{1}{2}} \theta^{\prime} \sin \theta^{\prime}
\\
& \qquad \qquad \qquad 
J_{1}(k\rho \sin \theta^{\prime})  e^{ik z \cos \theta^{\prime} } 
\label{Eq: circular integral func}
\, .
\end{split}
\end{align}

\subsection{Amplitude distribution}

We compute the amplitudes of electric field $\mathbf{E}$ and magnetic field $\mathbf{B}$ and evaluate the field-strength squared $F_{\mu \nu} F^{\mu \nu}$.
In the following, we focus only on the root mean square of each quantity.
That is,
\begin{align}
\begin{split}
\sqrt{\langle | \mathbf{E} (\mathbf{r}, t)|^{2} \rangle} 
&= \frac{1}{\sqrt{2}} \sqrt{ | \hat{\mathbf{E}} (\mathbf{r})|^{2}} 
\end{split}
\,, \\
\begin{split}
\sqrt{\langle | \mathbf{B} (\mathbf{r}, t)|^{2} \rangle} 
&= \frac{1}{\sqrt{2}} \sqrt{ | \hat{\mathbf{B}} (\mathbf{r})|^{2}} 
\end{split}
\,, \\
\begin{split}
\sqrt{\langle | F_{\mu \nu} F^{\mu \nu} (\mathbf{r}, t)|^{2} \rangle} 
&= \frac{2}{\sqrt{2}} \sqrt{ |  \hat{\mathbf{B}}^{2} (\mathbf{r}) - \hat{\mathbf{E}}^{2} (\mathbf{r}) |^{2}} 
\, .
\end{split}
\end{align}
Based on the experimental setup, we utilize the set of parameters in Table~\ref{Table: parameter}.
\begin{table}[htbp]
\centering
\begin{tabular}{ll}
\hline
Parameters &  \\
\hline
Creation beam frequency & 2.8~GHz \\
Inducing beam frequency& 1.6~GHz\\
Creation beam diameter & 6.02~m \\
Inducing beam diameter & 6.10~m  \\
Common focal length  & 30~m \\ 
\hline
\end{tabular}
\caption{
Experimental parameters for the stimulated radar collider setup, as listed in~\cite{Homma:2019rqb}.
}
\label{Table: parameter}
\end{table}

From the geometrical parameters, we compute the aperture angle $\theta_{f}$ (see also Fig.~\ref{Fig: setup}),
\begin{align}
\theta_{f} = 2 \arctan \left( \frac{\mbox{beam radius [m]}}{\mbox{focal length [m]}} \right)
\, .
\end{align}
Hereafter, we use the parameters of creation beam in evaluating the amplitudes of the electric and magnetic fields. 
Furthermore, we normalize the amplitude of the electric field for the incident beam $e_{0}=1$, thus $A=kf/2$, 
to see how the electric and magnetic fields are amplified in the focused beam.

\begin{figure}[htbp]
\centering
\includegraphics[width=0.4\textwidth]{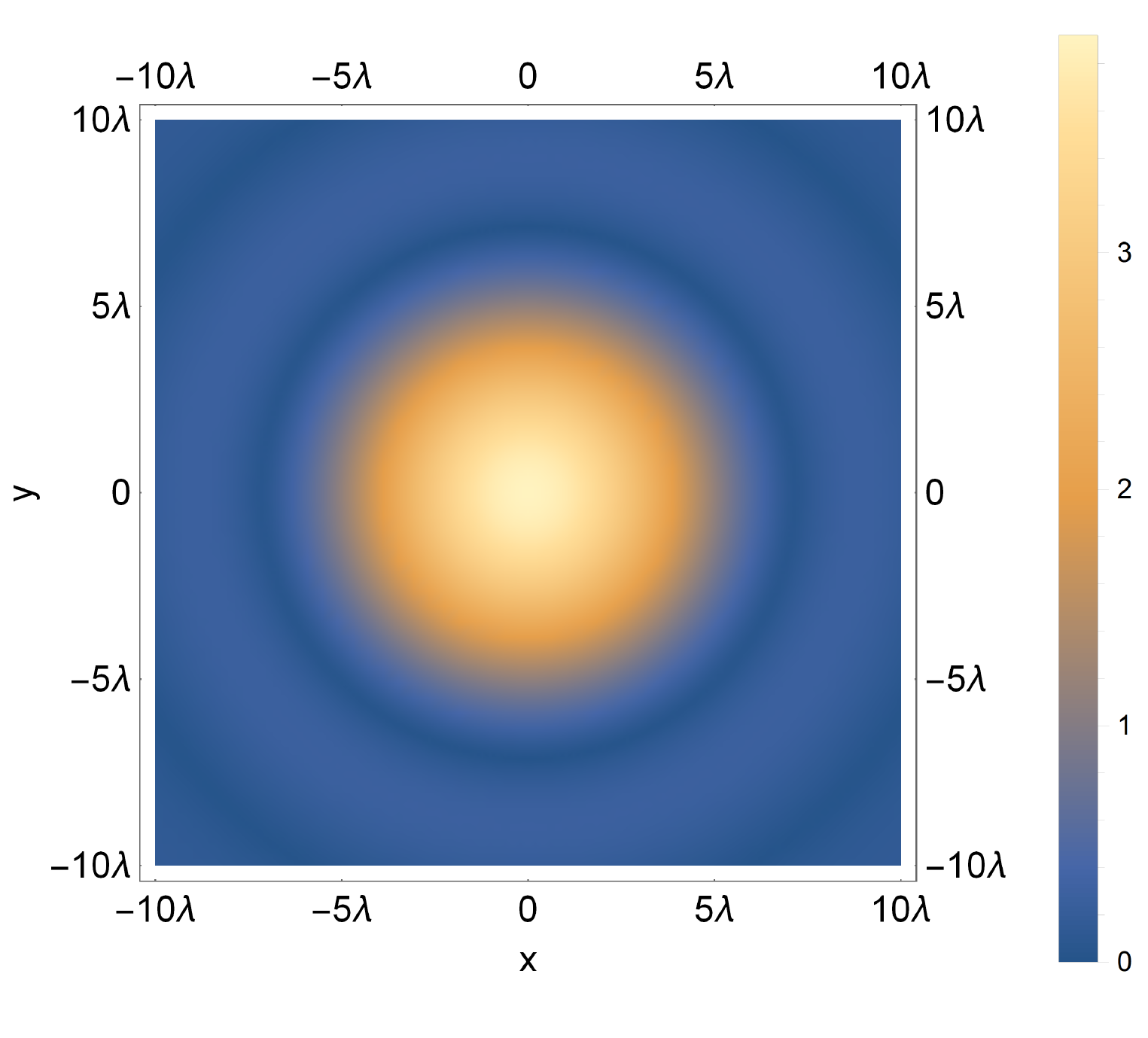}
\caption{
The root mean square of the electric field in linearly polarized beam, 
showing the amplitude distribution on the focal plane ($z=0$), where the center corresponds to the $z$-axis. 
}
\label{Fig: linear-electric}
\end{figure}

\begin{figure}[htbp]
\centering
\includegraphics[width=0.4\textwidth]{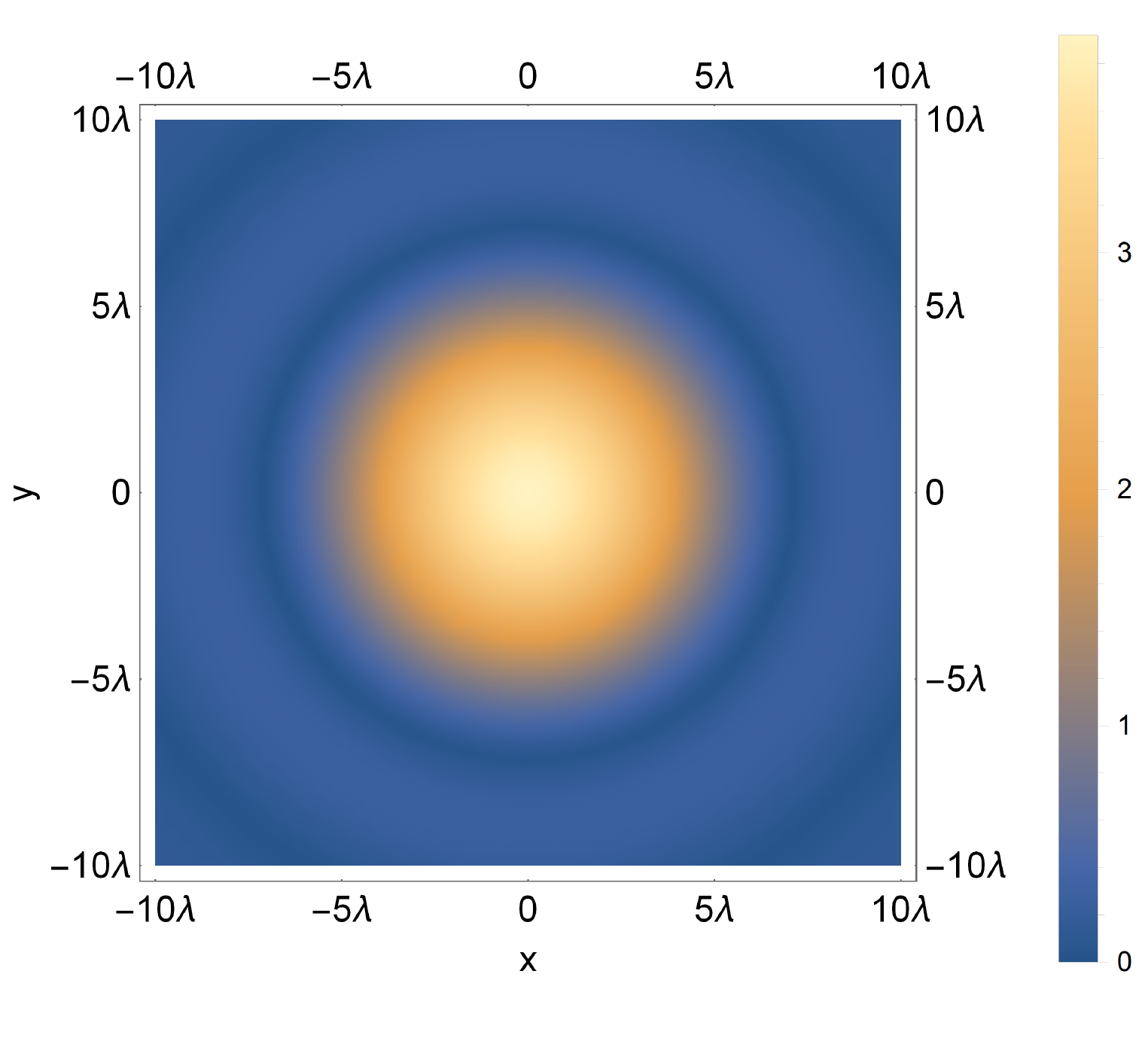}
\caption{
The same as Fig.~\ref{Fig: linear-electric}, but for 
the magnetic field in linearly polarized beam. 
}
\label{Fig: linear-magnetic}
\end{figure}

\begin{figure}[htbp]
\centering
\includegraphics[width=0.4\textwidth]{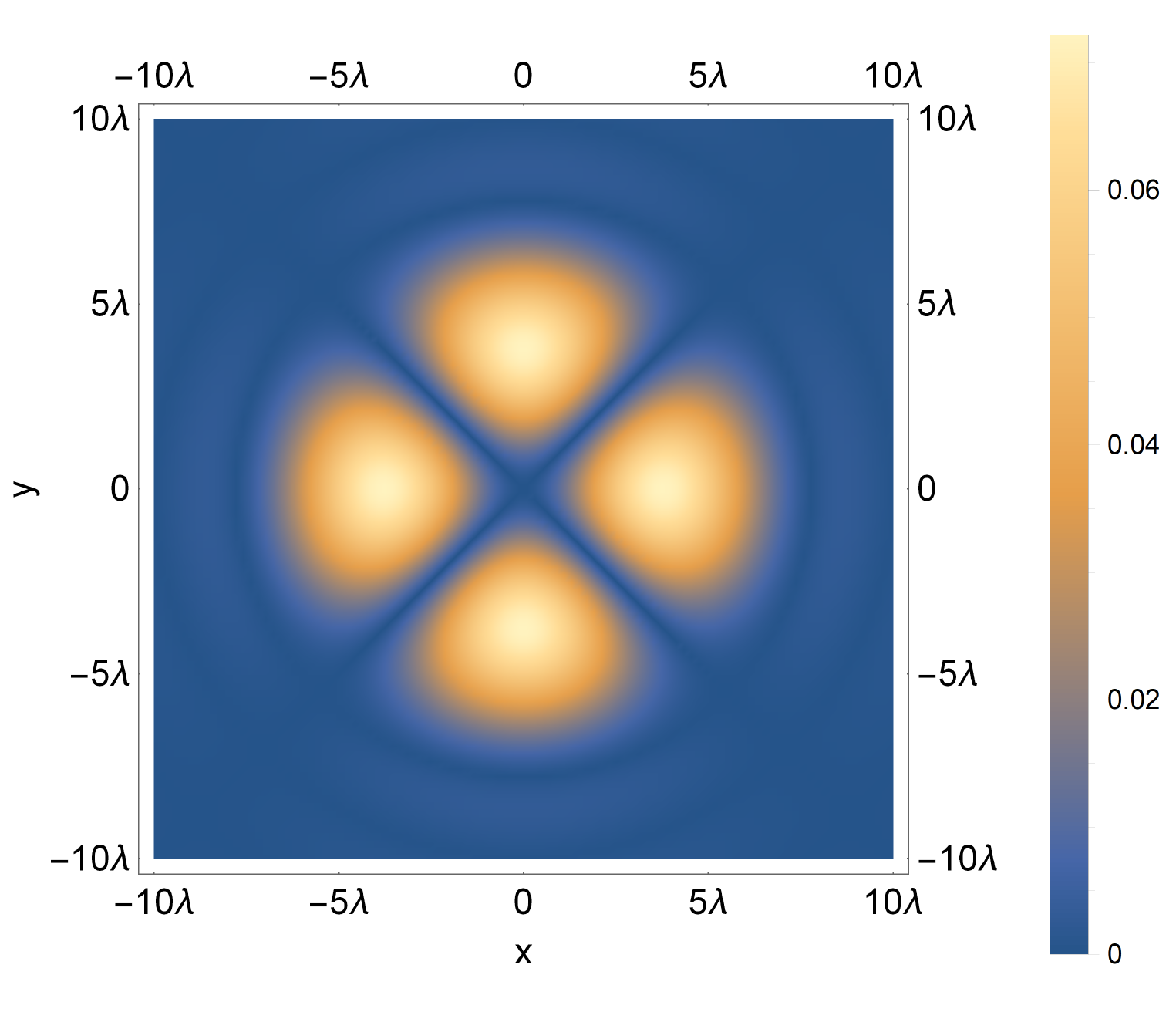}
\caption{
The same as Figs.~\ref{Fig: linear-electric} and~\ref{Fig: linear-magnetic}, 
but for 
the field-strength squared in linearly polarized beam.
}
\label{Fig: linear-Fmunu}
\end{figure}

First, we consider the case of focusing the linearly polarized beam.
We plot the amplitude distribution of the electric field, magnetic field, and field-strength squared on the focal plane ($z=0$)
in Figs~\ref{Fig: linear-electric}, \ref{Fig: linear-magnetic}, \ref{Fig: linear-Fmunu}.
Note that $e^{ik z \cos \theta^{\prime} }$ in Eq.~\eqref{Eq: linear integral func} vanishes on the focal plane. 
In each plot, the $(x, y)$ coordinate is in the unit of the wavelength $\lambda$
because the $\rho(x, y)$-dependence originates from $J_{n}(k \rho  \sin \theta^{\prime})$ and $k \rho = 2\pi \rho/\lambda$.
One finds that the amplitude distributions of the electric and magnetic fields are elliptical for the linearly polarized beam.
The plot of the amplitude distribution shows that the field-strength squared vanishes on $z$-axis ($\rho=0$) of the focal plane for the linearly polarized beam.

\begin{figure}[htbp]
\centering
\includegraphics[width=0.4\textwidth]{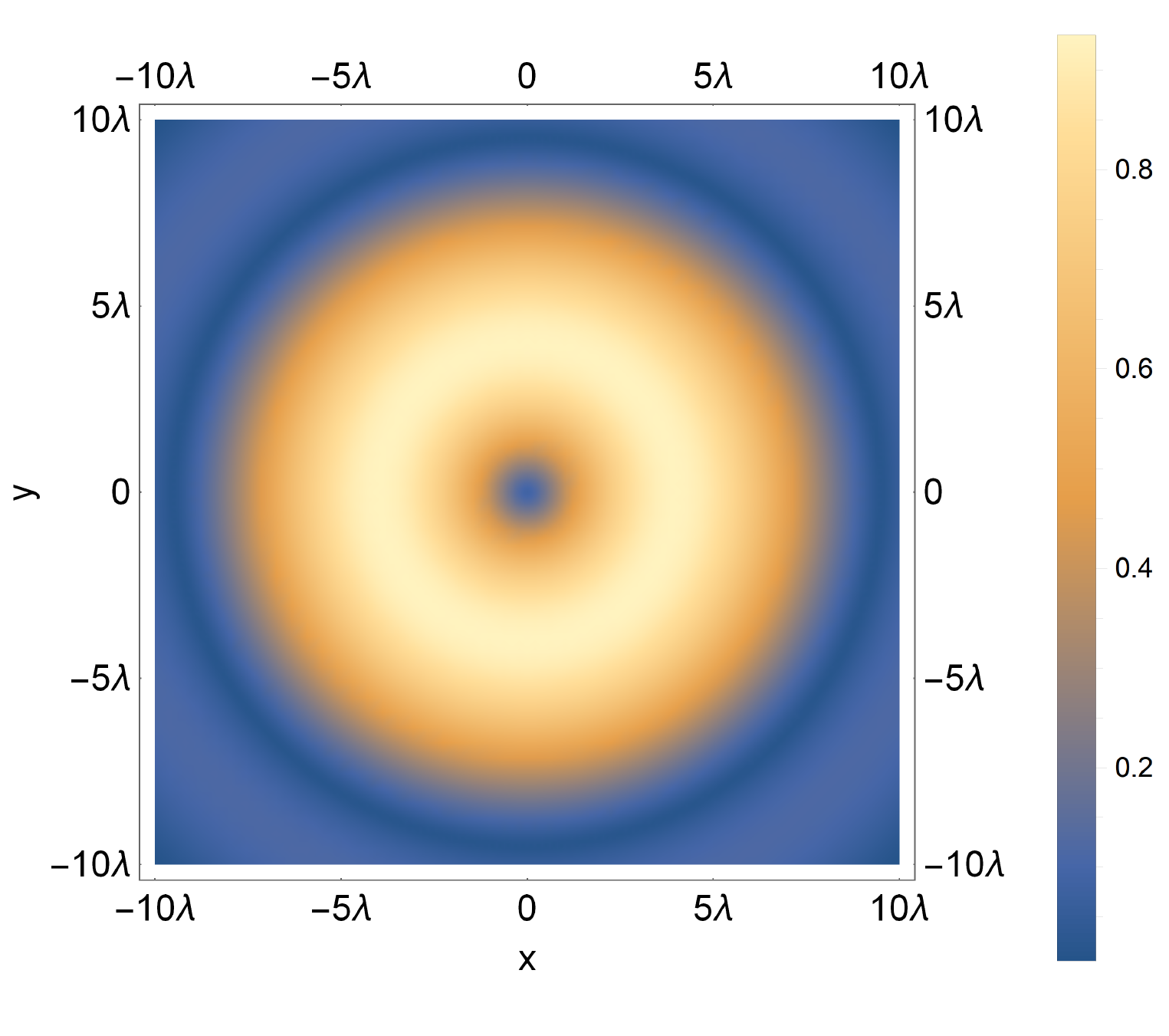}
\caption{
The root mean square of the electric field in circularly polarized beam. 
}
\label{Fig: circular-electric}
\end{figure}

\begin{figure}[htbp]
\centering
\includegraphics[width=0.4\textwidth]{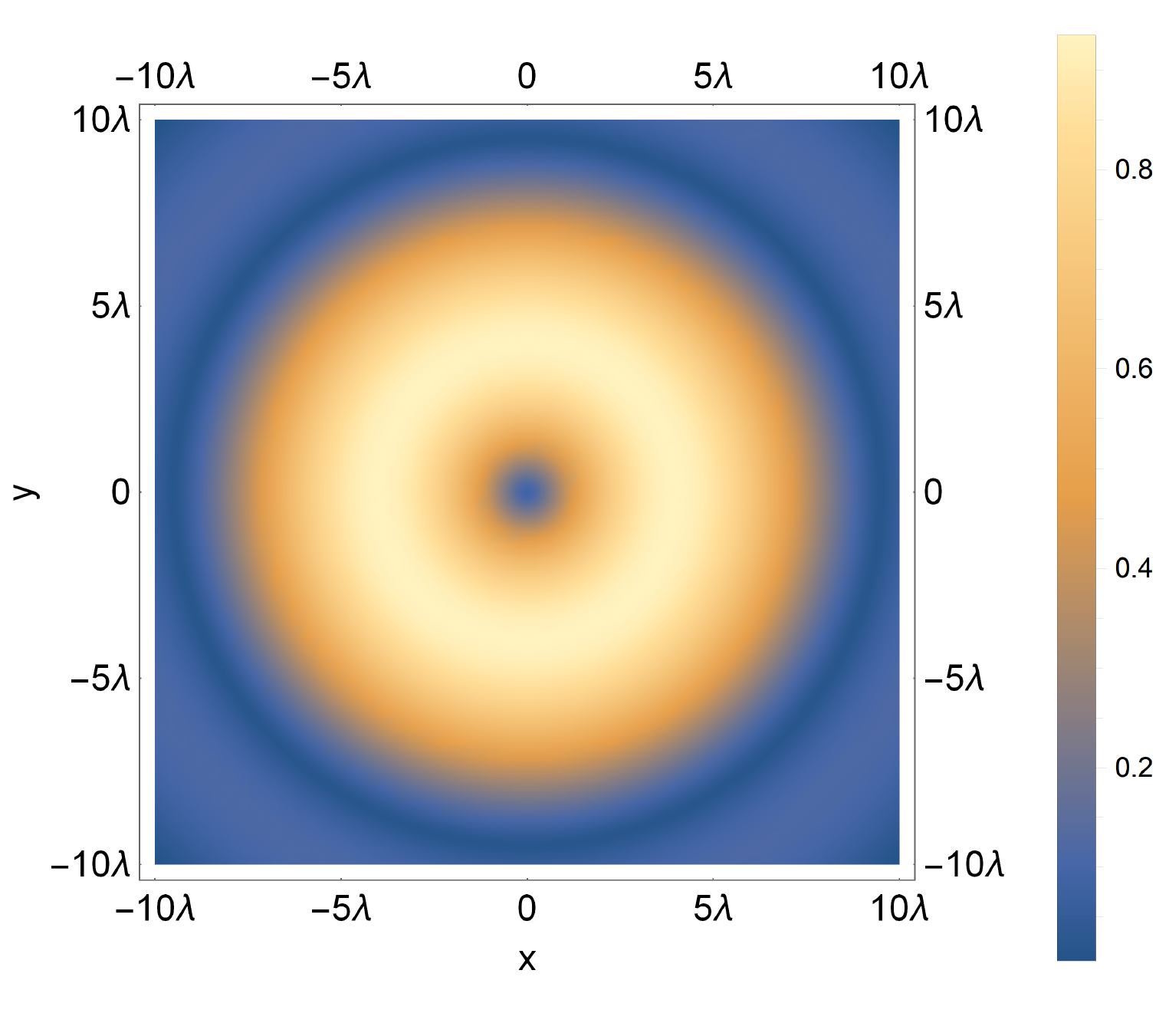}
\caption{
The root mean square of the magnetic field in circularly polarized beam. 
}
\label{Fig: circular-magnetic}
\end{figure}

Next, we consider the case of focusing the circularly polarized beam.
We plot the amplitude distribution of the electric field and magnetic field in Figs~\ref{Fig: circular-electric}, \ref{Fig: circular-magnetic}.
Equation~\eqref{Eq: circular integral func} shows that $U_{0}$ and $U_{2}$ vanishes when $\rho \rightarrow 0$, 
and only $\hat{E}_{z}$ and $\hat{B}_{z}$ on $z$-axis ($\rho=0$) contributes to the total amplitudes.
From Eq.~\eqref{Eq: circular elemag},
one finds that the electric and magnetic fields have the same amplitude at each point on the focal plane for the circularly polarized beam ($\varphi_{0} = \pi/4$).
Therefore, the field-strength square vanishes on the focal plane for the circularly polarized beam. 

\section{Dark Energy Models of $F(R)$ Gravity}
We briefly review two cosmological models of $F(R)$ gravity: 
Starobinsky~\cite{Starobinsky:2007hu} model and Hu-Sawicki model~\cite{Hu:2007nk}, 
which have been widely used in the study on late-time cosmology.
First, the Starobinsky model is describe by the following $F(R)$ function: 
\begin{align}
    F_{\mathrm{S}}(R) = R -\mu_{\mathrm{S}} R_{\mathrm{S}} \left[ 1 - \left( 1+\frac{R^{2}}{R^{2}_{\mathrm{S}}} \right)^{-l} \right]
    \, .
\end{align}
$R_{\mathrm{S}}$ represents a background curvature of the DE vacuum,
$\mu_{\mathrm{S}}$ and $l$ are positive parameters.
In the large-curvature limit $R\gg R_{S}$, the Starobinsky model approximates to
\begin{align}
    F_S(R) &\approx R -\mu_{\mathrm{S}} R_{\mathrm{S}} \left[ 1 - \left( \frac{R}{R_{\mathrm{S}}} \right)^{-2l} \right]
    \nonumber \\
    &=  R -\mu_{\mathrm{S}} R_{\mathrm{S}}+\mu_{\mathrm{S}} R_{\mathrm{S}} \left( \frac{R}{R_{\mathrm{S}}} \right)^{-2l} 
    \, .
\end{align}
The second term in the second line can be considered as the cosmological constant, $\mu_{S} R_{S} = 2 \Lambda$.

Second, the $F(R)$ function for the Hu-Sawicki model is: 
\begin{align}
    F_{\mathrm{HS}}(R) = R 
    - R_{\mathrm{HS}} \frac{c_{1} \left( \frac{R}{R_{\mathrm{HS}}} \right)^{m}}
    {1 + c_{2} \left( \frac{R}{R_{\mathrm{HS}}} \right)^{m}}
    \, ,
\end{align}
$R_{HS}$ represents the background curvature of the DE vacuum in this model,
$c_{1}$, $c_{2}$, and $m$ are positive parameters.
Here, we consider to change the parametrization for this
$F_{\rm HS}(R)$ as follows:
\begin{align}
    F_{\mathrm{HS}}(R) = R 
    - \mu_{\mathrm{HS}} \tilde{R}_{\mathrm{HS}} \frac{ \left( \frac{R}{\tilde{R}_{\mathrm{HS}}} \right)^{m}}
    {1 +\left( \frac{R}{\tilde{R}_{\mathrm{HS}}}\right)^{m}}
    \, ,
\end{align}
where
$\mu_{HS} = \frac{c_{1}}{c^{1-1/m}_{2}}$ and $\tilde{R}_{\mathrm{HS}}=\frac{R_{\mathrm{HS}}}{c^{1/m}_{2}}$.
In the large-curvature limit $R\gg \tilde{R}_{\mathrm{HS}}$, the Hu-Sawicki model approximates to
\begin{align}
    F_{\mathrm{HS}} (R) &\approx R -\mu_{HS} \tilde{R}_{HS} \left[ 1 - \left( \frac{R}{\tilde{R}_{HS}} \right)^{-m} \right]
    \nonumber \\
    &=  R -\mu_{\mathrm{HS}} \tilde{R}_{\mathrm{HS}} + \mu_{\mathrm{HS}} \tilde{R}_{\mathrm{HS}} \left( \frac{R}{\tilde{R}_{\mathrm{HS}}} \right)^{-m} 
    \, .
\end{align}
In a way similar to the case of the Starobinsky model, 
the second term in the second line can be considered as the cosmological constant,  
$\mu_{S} \tilde{R}_{S} = \frac{c_{1}}{c_{2}} R_{\mathrm{HS}} = 2 \Lambda$.

Therefore, those two models are reduced to the same form of $F(R)$ gravity in the large-curvature limit as follows:
\begin{align}
    F(R) = R -\lambda R_{c} \left[ 1 - \left( \frac{R}{R_{c}} \right)^{-2n} \right]
\end{align}
Correspondence of parameters to those in Starobinsky and Hu-Sawicki models reads 
\begin{align}
    R_{c} &= R_{S} \, , \ \lambda = \mu_{S} \, , \ n=l
    \\
   R_{c} &= \frac{R_{HS}}{c^{1/m}_{2}}\, , \ \lambda = \frac{c_{1}}{c^{1-1/m}_{2}} \, , \ n = \frac{m}{2}
   \, .
\end{align}
Note that when we take the limit $\mu \rightarrow \infty$ and $R_{c} \rightarrow 0$ with fixing $\mu R_{c}$,
we obtain the general relativity with the cosmological constant, that is, $\Lambda$-CDM model.

\section{Evaluation of Gas in Chamber}

The specific gas constants for typical residual gases are shown in  Table~\ref{Tab:gas}.
\begin{table}[htbp]
\centering
\begin{tabular}{lll}
\hline
Gas & Molar mass [kg/mol] & $\mathcal{R}$ \\
\hline
Hydrogen ($\mathrm{H_2}$) & $2.0159\times 10^{-3}$ & 4124 \\
Water vapor ($\mathrm{H_2 O}$) & $18.015\times 10^{-3}$ & 461.5 \\
Nitrogen ($\mathrm{N_2}$) & $28.013\times 10^{-3}$ & 296.8 \\
Carbon Monoxide ($\mathrm{CO}$) & $28.010\times 10^{-3}$ & 296.8 \\
Dry air (mixture) & $28.965 \times 10^{-3}$ & 287.1 \\
Carbon Dioxide ($\mathrm{CO_2}$) & $44.010 \times 10^{-3}$ & 188.9 \\
\hline
\end{tabular}
\caption{
Specific gas constant for several gas species.
}
\label{Tab:gas}
\end{table}

As an illustration, we consider the dry air. 
For the room temperature $T=300~[\mathrm{K}]$ and the pressure $P=10^{-7}~[\mathrm{Pa}]$,
we find $\rho =1.16 \times 10^{-12}~[\mathrm{kg/m^{3}}]$.
Note that $1~[\mathrm{Pa}] = 4.80 \times 10^{-2}~[\mathrm{eV^{4}}]$ and $1~[\mathrm{kg/m^{3}}] = 4.31 \times 10^{15}~[\mathrm{eV^{4}}]$,
and thus, $P = 4.80 \times 10^{-9}~[\mathrm{eV^{4}}]$ and $\rho =5.00 \times 10^{3}~[\mathrm{eV^{4}}]$.
Then, we can ignore the pressure in evaluating the trace of the energy-momentum tensor, 
thus, $T^{\mu}_{\ \mu (\mathrm{gas})} \approx - \rho$.  

\bibliographystyle{apsrev4-1}
\bibliography{reference}

\end{document}